\documentclass[11pt,a4paper]{article}
\pdfoutput=1
\usepackage{jcappub}

\usepackage[utf8]{inputenc}
\usepackage[T1]{fontenc}
\usepackage{graphicx,amssymb,amsmath,wasysym}
\usepackage{color}

\newcommand{\mh}{m_h}
\newcommand{\ms}{m_S}
\newcommand{\ls}{\lambda_S}
\newcommand{\lhs}{\lambda_{HS}}
\newcommand{\som}{\sigma_{SS}/\ms}

\newcommand{\neqq}{{n'}^\text{eq}}
\newcommand{\xpf}{x'_\text{FO}}
\newcommand{\xf}{x_\text{FO}}

\begin{document}

\title{$\boldsymbol{\mathbb{Z}_2}$ SIMP Dark Matter}

\author[1]{Nicolás Bernal}
\emailAdd{nicolas@ift.unesp.br}

\author[2]{and Xiaoyong Chu}
\emailAdd{xchu@ictp.it}

\affiliation[1]{ICTP South American Institute for Fundamental Research\\
                Instituto de Física Teórica, Universidade Estadual Paulista, São Paulo, Brazil}

\affiliation[2]{ICTP International Centre for Theoretical Physics\\
Strada Costiera 11, 34014 Trieste, Italy}

\abstract{
Dark matter with strong self-interactions provides a compelling solution to several small-scale structure puzzles.  
Under the assumption that the coupling between dark matter and the Standard Model particles is suppressed, such strongly interacting massive particles (SIMPs) allow for a successful thermal freeze-out through $N$-to-$N'$ processes, where $N$ dark matter particles annihilate to $N'$ of them.
In the most common scenarios, where dark matter stability is guaranteed by a $\mathbb{Z}_2$ symmetry, the seemingly leading annihilating channel, i.e.~3-to-2 process, is forbidden, so the 4-to-2 one dominate the production of the dark matter relic density.  Moreover, cosmological observations require that the dark matter sector is colder than the thermal bath of Standard Model particles, a condition that can be dynamically generated via a small portal between dark matter and Standard Model particles, à la freeze-in. This scenario is exemplified in the context of the Singlet Scalar dark matter model.
}

\maketitle

\section{Introduction}\label{sec:introduction}

One of the major tasks in particle physics is to accommodate dark matter (DM)~\cite{Ade:2015xua} into the Standard Model (SM)  and its extensions. The fact that so far only the gravitational interactions of DM have been confirmed makes this challenging. Among the various strategies used to hunt for DM direct interactions, progress in precision cosmology allows, for the first time,  to look for possible signatures of non-gravitational self-interactions among DM particles in the sky. 
The expectation is that strong enough DM self-interactions, which are mostly likely to be short-ranged,
would leave observable imprints on small scales during the growth of cosmic structures.
In fact, some small-scale puzzles have been reported, such as the so-called `cusp vs. core'~\cite{Flores:1994gz,Moore:1994yx,Oh:2010mc,Walker:2011zu} and the `too-big-to-fail'~\cite{BoylanKolchin:2011de,Garrison-Kimmel:2014vqa} problems, challenging the standard collisionless cold DM paradigm. Both problems can be addressed by strong DM self-interactions~\cite{Spergel:1999mh,Wandelt:2000ad,Vogelsberger:2012ku,Rocha:2012jg,Peter:2012jh,Zavala:2012us,Kaplinghat:2015aga}, although astrophysical solutions also exist~\cite{MacLow:1998wv,Governato:2009bg, Silk:2010aw, VeraCiro:2012na}. More recently, a non-vanishing offset between DM and star mass distribution in the cluster Abell 3827 has been reported, also  pointing towards to the existence of strong DM self-interactions~\cite{Massey:2015dkw, Kahlhoefer:2015vua}. 

Strong self-interactions among DM particles would have a significant impact on other aspects of DM phenomenology. For instance, frequent self-scatterings among DM particles modify their phase-space distribution in the Local Group, as well as inside astrophysical objects, leading to novel signatures/interpretations in DM direct and indirect searches~\cite{Ganguly:2006ki, Zentner:2009is, Vogelsberger:2012sa, Ng:2013xha,DelNobile:2015uua}.
More importantly, to achieve sufficient DM self-interactions without contradicting cosmological observations, new mechanisms in DM modeling need to be introduced.\footnote{For concrete examples see Refs.~\cite{Carlson:1992fn, Wandelt:2000ad, McDonald:2001vt, Gudnason:2006yj, Khlopov:2008ki,Kamionkowski:2008gj, Kaplan:2009de, Feng:2009hw, Tulin:2012wi, Schutz:2014nka, Hochberg:2014dra, Kouvaris:2014uoa}, and for a recent review  see Ref.~\cite{Kusenko:2013saa}.} The DM self-interactions strength required to alleviate the small-scale structure problems is indeed more than ten orders of magnitude larger than the weak interactions. In this context, the existence of strong DM self-interactions also influences the DM production mechanisms.
In particular, the DM thermal freeze-out does not have to be induced by the usual annihilations of two DM particles into a couple of lighter states. 

Alternatively, one can consider a framework where the freeze-out proceeds via $N$-to-$N'$ number-changing processes, where $N$ DM particles annihilate into $N'$ of them (with $N>N'\geq 2$), first studied in Ref.~\cite{Carlson:1992fn} and recently named the `SIMP paradigm'~\cite{Hochberg:2014dra}.
One potential problem of this mechanism is that when the $N$-to-$N'$ annihilations are effective, DM reheats itself, modifying significantly the formation of structures~\cite{deLaix:1995vi}.
Several solutions exist: One can assume either kinetic equilibrium between the DM and the visible sector~\cite{Hochberg:2014dra,Choi:2015bya}, or an enlarged dark sector containing new particles that are relativistic at the moment of the freeze-out~\cite{Bernal:2015bla}.
A different approach, adopted in this work, is to assume that the two sectors never reach kinetic equilibrium with each other, and that the DM temperature $T'$ is much smaller than the SM temperature $T$~\cite{Bernal:2015}.
This can be easily achieved through a very weak portal between DM and SM particles, as shown later.

The most studied cases of the $N$-to-$N'$ processes correspond to 3-to-2~\cite{Carlson:1992fn,Hochberg:2014dra}, because it is typically the dominant one.
However, 3-to-2 annihilations, necessarily induced by interaction vertices with an odd number of DM particles, are forbidden  in the most common models where  the DM stability is guaranteed by a ${\mathbb Z}_2$ symmetry (e.g. $R$-parity in SUSY or $K$-parity in Kaluza-Klein scenarios).
To allow for 3-to-2 annihilations, one has to assume that DM is protected by a different symmetry such as a ${\mathbb Z}_3$~\cite{Ko:2014nha, Bernal:2015bla,Choi:2015bya}, or consider models where the DM stability emerges as a result of the DM dynamics~\cite{Yamanaka:2014pva,Hochberg:2014kqa,Lee:2015gsa,Hansen:2015yaa,Bernal:2015}.
If DM is stabilized by a ${\mathbb Z}_2$ symmetry, the 4-to-2 reactions would be the ones giving rise to the DM relic abundance, while the 3-to-2 annihilations are forbidden.
In this work we study such SIMP DM candidates, and illustrate the results in the Singlet Scalar DM model~\cite{McDonald:1993ex,Burgess:2000yq} as an example.

In Section~\ref{sec:ssdm} we briefly review the Singlet Scalar DM model.
Section~\ref{sec:dmsi} is dedicated to the conditions for this model to alleviate the small scale problems with DM self-interactions.
Then we discuss in Section~\ref{sec:T} the DM production mechanism compatible with the strong self-interactions: 4-to-2 DM annihilations, with DM particles colder than the SM.
A small Higgs portal can naturally generate such a difference in their temperatures, as shown in Section~\ref{sec:lhs}.
In Section~\ref{sec:fermionicDM} we briefly discuss ways of mitigating the strong constraints on DM self-interactions from cluster observations, that could allow for lighter DM, before concluding in Section~\ref{sec:conclusions}.

\section{Singlet Scalar $\boldsymbol{\mathbb{Z}_2}$  Dark Matter}
\label{sec:ssdm}

The singlet scalar model~\cite{McDonald:1993ex,Burgess:2000yq} is one of the minimal extensions of the SM that can provide a viable DM candidate. In addition to the SM framework, this model contains a scalar sector which is only composed by a real scalar $S$. This particle is a singlet under the SM gauge group, but odd under a $\mathbb{Z}_2$ symmetry, which guarantees its stability. 
The most general renormalizable scalar potential is given by
\begin{equation}
V=\mu_H^2\,|H|^2+\lambda_H\,|H|^4+\mu_S^2\,S^2+\ls\,S^4+\lhs\,|H|^2\,S^2\,,
\end{equation}
where $H$ is the SM Higgs doublet.
We require that the Higgs gets a non-vanishing vacuum expectation value, $v_H = 246$~GeV, while the singlet does not, $\langle S\rangle = 0$, to ensure the stability of the DM candidate.
At tree level, the singlet mass is $\ms^2=2\,\mu_S^2+\lhs\,v_H^2$.
The phenomenology of this model is completely determined by three parameters: the DM mass $\ms$, the Higgs portal $\lhs$ and the quartic coupling $\ls$. Note that the role of the DM self-coupling $\ls$ was typically disregarded in previous studies.

There has been a large amount of research on the singlet scalar DM model, most of them focused on the WIMP scenario, where the singlet $S$ mixes relatively strongly with the Higgs and undergoes a thermal freeze-out. This scenario has been highly constrained by collider searches~\cite{Barger:2007im,Djouadi:2011aa,Djouadi:2012zc,Damgaard:2013kva,No:2013wsa,Robens:2015gla}, DM direct detection~\cite{He:2009yd,Baek:2014jga, Feng:2014vea, Queiroz:2014yna, Han:2015hda} and indirect detection~\cite{Yaguna:2008hd, Goudelis:2009zz, Profumo:2010kp,Cline:2013gha,Urbano:2014hda, Duerr:2015mva,Duerr:2015aka}. In contrast, scenarios with a very suppressed Higgs portal are much less constrained, and could also lead to a vast phenomenology, such as the freeze-in mechanism~\cite{McDonald:2001vt, Hall:2009bx, Yaguna:2011qn,Campbell:2015fra,Kang:2015aqa}.  The paradigm studied in this work belongs to the latter category.

\vspace{.1cm}

For the phenomenological study in the following sections, the Lagrangian of the model is implemented in {\tt SARAH}~\cite{Staub:2008uz,Staub:2013tta} and {\tt CalcHEP}~\cite{Belyaev:2012qa}.

\section{Dark Matter Self-Interactions}
\label{sec:dmsi}

\begin{figure}[t!]
\begin{center}
\includegraphics[height=2.8cm]{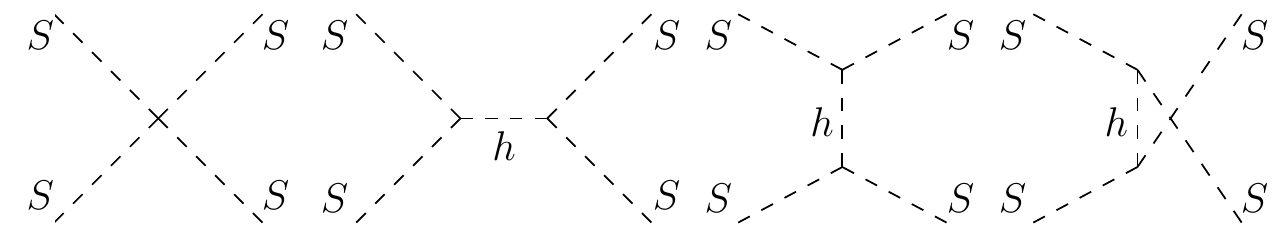}
\end{center}
\caption{ {{Feynman diagrams for DM self-interactions.}} In the limit of either $m_S \ll m_h$ or small $\lambda_{HS}$, the contract interaction (first diagram) dominates.
}
\label{fig:selfDM}
\end{figure}

In the singlet scalar model, self-interactions occur via the contact interaction and the $s$-, $t$- and $u$-channel exchange of a Higgs boson (Fig.~\ref{fig:selfDM}).
If DM is light (i.e. $\ms\ll\mh$), the first process typically dominates.
In this case, the corresponding cross-section at low redshifts depends only mildly on the non-relativistic DM velocity.
In the limit of a small Higgs portal and low velocity, the ratio of the DM self-interaction cross section over the DM mass is given by

\begin{equation}\label{SS:sigma}
\frac{\sigma_{SS}}{\ms}=\frac{9}{8\,\pi}\frac{\ls^2}{\ms^3}\left[1-\frac{\lhs^2\,v_H^2}{3\,\ls\,\mh^2}\frac{3\,\mh^2-8\,\ms^2}{\mh^2-4\,\ms^2}\right]^2
\sim\frac{9}{8\,\pi}\,\frac{\ls^2}{\ms^3}\,.
\end{equation}

In order to smooth the inner region of DM halos at scales of few kpc, as required by the `cusp vs. core' problem, cosmological simulations show that the DM self-interaction needs to lay in the range $0.1\lesssim\som\lesssim10$ cm$^2/$g, while a smaller value is enough to solve the `too-big-to-fail' problem~\cite{Vogelsberger:2012ku,Rocha:2012jg,Peter:2012jh,Zavala:2012us,Vogelsberger:2014pda,Elbert:2014bma,Kaplinghat:2015aga}.
On the other hand, too strong self-interactions could cause a collapse in the core of DM halos~\cite{Vogelsberger:2012ku}.
Concretely, Bullet Cluster measurements point towards DM self-interactions below $1.25$ cm$^2$/g at $68\%$~CL~\cite{Markevitch:2003at,Clowe:2003tk,Randall:2007ph}. Moreover, recent observational data on cluster collisions have led to a more stringent bound: $\som<0.47$~cm$^2$/g at $95\%$~CL~\cite{Harvey:2015hha}.

Given the large uncertainties in cosmological simulations, we will consider $0.1\lesssim\som\lesssim1$ cm$^2/$g as the parameter range of interest throughout the paper. Having in mind Eq.~\eqref{SS:sigma} together with the condition of perturbativity $\ls<4\pi$, the singlet scalar model naturally leads to a DM mass in the sub-GeV region: $\ms\lesssim 200$~MeV. Such a light DM allows the Higgs to decay into a pair of DM particles. Therefore, both the current limits on the invisible Higgs branching ratio (BR$_\text{inv}\lesssim 20\%$~\cite{Bechtle:2014ewa}) and the Higgs total decay width ($\Gamma_h^\text{tot}\lesssim 22~\text{MeV}$~\cite{Khachatryan:2014iha}) strongly constrain the strength of the Higgs portal, resulting in $\lhs<7\cdot 10^{-3}$.

To summarize, the solution to the small scale problems implies a sub-GeV DM with strong self-coupling $\ls\sim\mathcal{O}(1)$.
On top of that, to avoid a too large invisible decay of the Higgs in conflict with the LHC data, a suppressed Higgs portal is also required.

\section{Dark Freeze-out with $\boldsymbol{T' < T}$}
\label{sec:T}

In the singlet scalar model, the DM relic abundance can not be generated by the usual pair-annihilating freeze-out, if one wants to solve at the same time the small scale structure problems.  In fact, for $\ms\lesssim 200$~MeV, the kinematically-allowed final states consist of only pions and light fermion pairs (e.g. muons, electrons, etc.). The thermally averaged cross-section in the non-relativistic approximation is
\begin{equation}
\langle\sigma v\rangle_{SS\to f\bar f}\sim\frac{\lhs^2}{\pi}\frac{m_f^2}{\mh^4}\,,
\end{equation}
which is very suppressed because of the small fermion mass $m_f$ compared to the `heavy' Higgs. For instance, $\lhs \sim 10^{-2}$ and $m_f = m_\mu$ give $\langle\sigma v\rangle_{SS\to f\bar f} \sim 10^{-32}$~cm$^3$/s, far below the canonical value of few $10^{-26}$~cm$^3$/s~\cite{Steigman:2012nb}, and thus leads to an overclosed Universe. Hence, DM can not be a vanilla thermal WIMP annihilating into a pair of SM particles. Another mechanism for generating the DM relic abundance is therefore needed.

\begin{figure}[t]
\centering
\includegraphics[height=3.0cm]{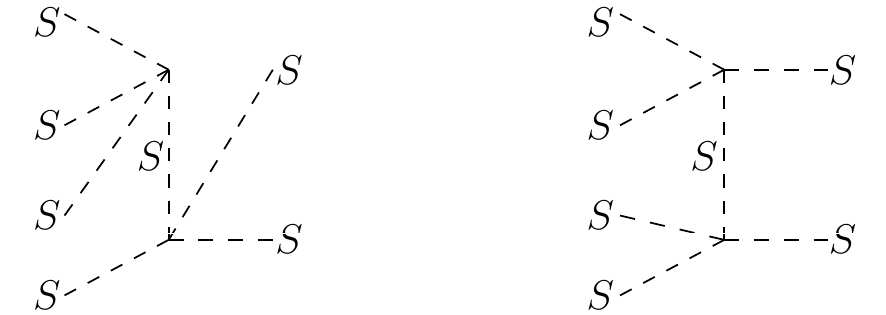}
\caption{{{Tree level diagrams for the process ${SSSS\leftrightarrow SS}$, in the limit of small ${\lhs}$.}}
}
\label{xx-xxxx}
\end{figure}

Under these conditions, a freeze-out via $N$-to-$N'$ reactions  emerges as the dominant DM production mechanism.
As DM only annihilates within the dark sector, it is also referred to as `dark freeze-out'~\cite{Feng:2008mu,Ackerman:mha}.\footnote{While we are mostly concerned with the dark freeze-out scenario here, Ref.~\cite{Bernal:2015} provides a general discussion on various production mechanisms of self-interacting DM.}
In practice, this scenario can give rise to both large DM self-interactions and the proper DM relic abundance, while being consistent with all experimental constraints. In the singlet scalar model, it happens dominantly via 4-to-2 annihilating channels, as shown in Fig.~\ref{xx-xxxx}.
This is because the ${\mathbb Z}_2$ symmetry forbids the 3-to-2 interactions in the model.
Higher-order processes, such as the 5-to-3 or the 6-to-2 annihilations, are sub-dominant and thus not considered hereafter.
The 4-to-2 annihilations have been considered in Ref.~\cite{Hochberg:2014dra} assuming that the visible and dark sectors share the same temperature. They concluded that the dark freeze-out requires a too light DM, in tension with cosmological observations.  Besides, Refs.~\cite{Campbell:2015fra,Kang:2015aqa} studied the production of self-interacting singlet scalar DM via the freeze-in without taking into account the effect of 4-to-2 DM annihilations, which is actually important as shown in Section~\ref{sec:lhs}.

\vspace{.2cm}

\begin{figure}[t]
\centering
\includegraphics[width=0.58\textwidth]{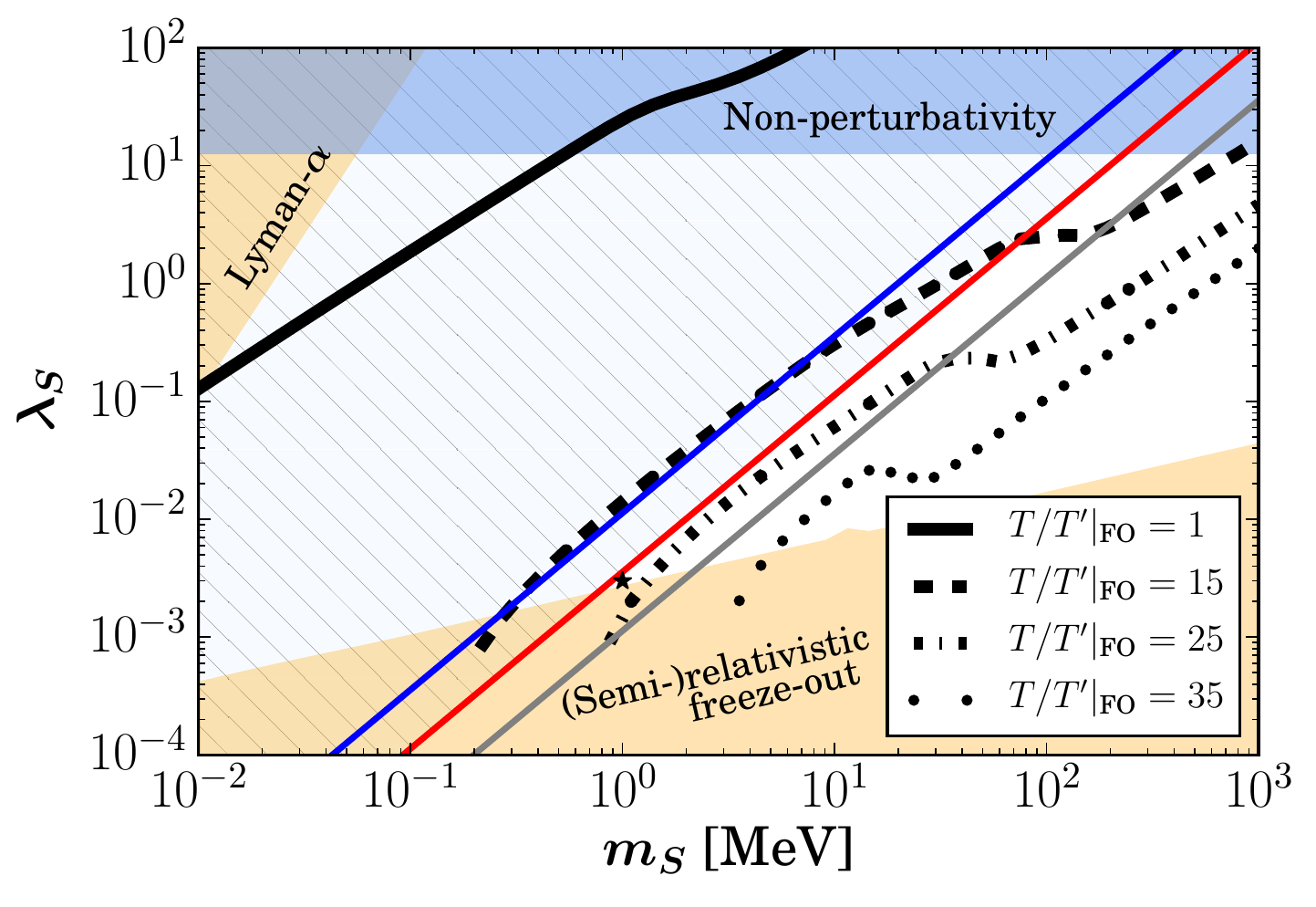}
\vspace{-.4cm}
\caption{Regions on the plane of $\ls$-$m_S$ that yield the observed DM relic density. 
The four thick black lines correspond to the 4-to-2 DM annihilations, for various temperature ratios at the moment of the freeze-out.
The gray, red and blue lines correspond to $\som=0.1$, 1 and 10 cm$^2/$g, respectively,  whereas the region excluded by cluster observations is hatched. The area where $\lambda_S \ge 4\pi$ is shown in light-blue. The two orange bands correspond to a (semi-)relativistic freeze-out and the Lyman-$\alpha$ bound,
respectively. The star corresponds to the parameter point used for Fig.~\ref{thermal:dark}. See text for details.
}
\label{ms-ls}
\end{figure}

In the case where the Higgs portal is very suppressed, the Boltzmann equation which describes the evolution of the DM number density $n'\equiv n'(T')$ is:
\begin{equation}
\frac{dn'}{dt}+3H(T)\,n'=- \langle\sigma v^3\rangle_{4\to 2}\left[n'^4-n'^2\,{n'}_\text{eq}^2(T')\right]\,,\label{Boltzmann:eqn}
\end{equation}
where $H(T)$ is the Hubble rate as a function of visible temperature $T$ and $n'_\text{eq}(T')$ represents the equilibrium number density of DM particles at a dark temperature $T'$.
In the non-relativistic limit, the 4-to-2  DM annihilation cross section is given by 
\begin{equation}
\langle\sigma v^3\rangle_{4\to 2}\sim\frac{27\sqrt{3}}{8\,\pi}\frac{\ls^4}{\ms^8}\,.
\end{equation}
In the Boltzmann equation~\eqref{Boltzmann:eqn}, a factor of 2, denoting that each $4\to2$ reaction removes two DM particles, has been absorbed into the definition of the cross section above.
This differential equation has been analytically solved in Appendix~\ref{Boltzman:FO}, under the assumption that the dark freeze-out happens instantaneously at a visible temperature $T_\text{FO}$ (or equivalently, $T'_\text{FO}$ for the dark sector), and the results are shown in  Fig.~\ref{ms-ls}. The solution is also based on the assumption that the dark freeze-out happens non-relativistically ($x'_\text{FO}\equiv\ms/T'_\text{FO}\gtrsim 3$), which corresponds to
\begin{equation}
\label{DFO:boundary}
\left.\frac{\Gamma_{4\to 2}(T')}{H(T)}\right|_{x'\sim 3} \gtrsim 1\,, 
\end{equation}
where $\Gamma_{ 4\to 2}=(n'_\text{eq})^3\,\langle\sigma v^3\rangle_{4\to 2}$.
Conversely, the orange lower band in Fig.~\ref{ms-ls} yields a (semi-) relativistic freeze-out (i.e. when Eq.~\eqref{DFO:boundary} is not satisfied), which will be addressed afterwards. 
We emphasize that this inequality requires that the DM particles were in chemical equilibrium between themselves before the freeze-out took place.

In Fig.~\ref{ms-ls}, the four black thick lines  show the coupling $\ls$ needed to reproduce the observed DM relic abundance through 4-to-2 annihilations. In the case where the dark and the visible sectors are in kinetic equilibrium (thick solid black line $T/T^\prime|_\text{FO} =1$), the perturbativity of the $\ls$ coupling implies a DM mass lighter than $\sim 500$~keV, in agreement with Ref.~\cite{Hochberg:2014dra}.
However, this case always produces a too large DM self-scattering.
This is shown  in Fig.~\ref{ms-ls} by the three straight diagonal lines, which correspond to the ratio $\som=0.1$, 1 and 10~cm$^2$/g (gray, red and blue lines, respectively), typically needed in order to solve the small-scale problems. The hatched area is disfavored by cluster observations (i.e. $\som>1$~cm$^2$/g).

The constraints from cluster observations become much less severe if the dark temperature $T'$ of DM particles is lower than the visible temperature $T$.
The reason is that in the case where $T/T^\prime|_\text{FO}\gg 1$, the equilibrium DM number density is suppressed by a factor $(T/T')^3$ compared to the case where $T=T'$. As a result, in order to obtain the observed DM abundance, an earlier freeze-out and thus a smaller annihilation rate are required. This corresponds to a smaller value for $\lambda_S$, leading to a weaker DM self-scattering more in line with the cluster bounds.
Fig.~\ref{ms-ls} illustrates this fact, by the three black lines (dashed, dotted-dashed, dotted) corresponding to  $T/T^\prime|_\text{FO} = 15$, 25 and 35, respectively.
From the figure one can see that for each parameter set ($\ms,\,\ls$) a ratio $T/T^\prime|_\text{FO}$ exists that gives rise simultaneously to the observed DM abundance and to the expected value of $\som$. The top blue region in Fig.~\ref{ms-ls} corresponds to the parameter space where non-perturbativity effects become important ($\ls > 4\pi$).  Moreover, the cosmological bound from Lyman-$\alpha$ correspond to the orange band in the top-left corner, where we conservatively take $T_\text{FO} \lesssim 1$~keV~\cite{Viel:2013apy}. 
\vspace{.1cm}

Let us recall that the entropies of the dark and the visible sectors are separately conserved after the moment when the two sectors kinetically decoupled from each other.
It is then useful to compute the entropy ratio between them, which is also a conserved quantity.
In particular, if the freeze-out happens non-relativistically, one has
\begin{equation}
\label{entropy:cor}
\xi \equiv \frac{s}{s'}  \sim \frac{T'_\text{FO}}{3.6~\text{eV}\,(1+ 2.5\,T'_\text{FO}/\ms) \cdot  \Omega_S h^2}\,,
\end{equation}
where $s$ and $s'$ are the entropy densities of the SM sector and the dark sector, respectively. This expression was first derived in Ref.~\cite{Carlson:1992fn}, and here we add a new term, $2.5~T'_\text{FO}/m_S$, to take into account second-order corrections.
Substituting $\Omega_S h^2 \sim 0.12$~\cite{Ade:2015xua} and the solution of $x'_\text{FO}$ in Appendix~\ref{Boltzman:FO} into the above equation, we estimate that the proper DM abundance requires 
\begin{equation}
\label{entropy:observed}
\xi  \sim  7\cdot 10^9 \, \left(\frac{\ms}{\text{GeV}}\right) \frac{1}{1 + 0.03 \left(\ls\,\frac{\text{GeV}}{\ms}\right)^{4/7}}\,.
\end{equation}
The entropy ratio is nearly proportional to the DM mass, up to a weak dependence on the self-coupling $\lambda_S$. Moreover, to match the observed DM relic abundance there is always a $\xi\gg 10^2$ for each DM mass. The model therefore never leads to observable extra radiation~\cite{Carlson:1992fn}, consistent with BBN/CMB bounds.

At last, we comment the opposite case where the freeze-out happens when the DM is (semi-)relativistic, corresponding to the lower band of Fig.~\ref{ms-ls}. Here the entropy ratio is obtained as
\begin{equation}
\label{entropy:cor2}
\xi  \sim \frac{m_S}{13.5\,\text{eV}  \cdot  \Omega_S h^2}\,,
\end{equation}
using an analogy with the hot freeze-out of SM neutrinos. The comparison between the previous two cases suggests that Eq.~\eqref{entropy:cor} can also give a good estimation for the (semi-) relativistic freeze-out by taking $T'_\text{FO}\sim \ms$.

\section{Generating $\boldsymbol{T' < T}$ via the Higgs Portal}
\label{sec:lhs}

In the previous section we studied the dark freeze-out of DM particles colder than the SM sector, without assuming any particular mechanism generating such a difference of temperatures.
This could happen even in the absence of the Higgs portal, provided that most DM particles were created at a much earlier time, for instance directly from the inflaton decay~\cite{Dev:2013yza,Kane:2015qea}, and then reached a chemical equilibrium within the dark sector. 

Now we explore the possibility that most of the DM particles were instead created from the SM sector through the Higgs portal. 
For a very small DM self-coupling, $\ls\lesssim 10^{-3}$, the freeze-in scenario takes place: once created, DM only redshifts with the expansion of the Universe, without interacting with other particles.
Therefore, it is similar to the case of the relativistic freeze-out above, except that now DM particles statistically carry an average kinetic energy similar to the one of the SM particles, i.e. of the order of $T$.
For light DM ($m_S \ll m_h/2$), the invisible decay of the Higgs dominates the energy transfer~\cite{McDonald:2001vt,Hall:2009bx, Yaguna:2011qn, Chu:2011be}, so it is straightforward to obtain
\begin{equation}
\Omega_S h^2  \sim 0.12\cdot \left(\frac{\lhs}{10^{-10}}\right)^2  \cdot \left(\frac{m_S}{1~\text{MeV}}\right)^2. \label{FI:lambdaM}
\end{equation}
Nevertheless, since the average kinetic energy of DM particles is given by $T$, Lyman-$\alpha$ forest surveys apply directly, requiring DM to be heavier than several keV~\cite{Viel:2013apy}.\footnote{Technically, Lyman-$\alpha$ bound still can not be applied strictly, but detailed calculations of DM momentum distribution function, together with reheating processes in SM sector, are beyond the scope of this work.}

\vspace{.1cm}

For higher values of $\lambda_S$ ($\ls\gtrsim 10^{-3}$), the situation is different. One would expect that if both $\ls$ and $\lhs$ are large enough, after a large amount of DM particles are produced from the SM sector, they can chemically thermalize with themselves: the `dark thermalization'~\cite{Chu:2013jja}. In this case the freeze-in scenario does not apply any more.  But the condition for obtaining the dark thermalization is not obvious.
To simplify the question, we assume that DM particles frequently scatter between themselves, reaching the kinetic equilibrium described by a dark temperature $T'$ and a non-vanishing chemical potential $\mu'$.  However, elastic scatterings do not change the DM number density, so one also needs to consider the leading DM number-changing channel, i.e. the 4$\leftrightarrow$2 processes, in order to reach chemical equilibrium.
The condition for the dark thermalization is therefore given by
\begin{equation}
\left.\frac{n'\,\langle\sigma v\rangle_{2\to 4}(T') }{H(T)}\right|_{x'\sim1/2}  \gtrsim 1\,,
\label{thermalize:2to4}
\end{equation}
evaluated just before the 2-to-4 processes become Boltzmann-suppressed.
Note that contrary to Eq.~\eqref{DFO:boundary}, we are taking here $n'$ and not $n'_\text{eq}$.
In practice, this condition is slightly looser than the one of Eq.~\eqref{DFO:boundary}, and they mostly coincide.

This condition for dark thermalization can be further verified by actually solving the full set of Boltzmann equations, including the Higgs portal interactions, for both the DM number density and its energy density.  One numerical solution is depicted in Fig.~\ref{thermal:dark} for a specific parameter set ($\ms=1$~MeV, $\ls= 4\cdot 10^{-3}$ and $\lhs=6\cdot 10^{-11}$), showing the evolution of both the DM abundance $Y_S~(\equiv n'/s)$ (left panel) and its dark temperature $T'$ (right panel).
The actual evolution is given by the two red lines, labeled $Y_S$ and $T'$, respectively.
It suggests that even for $\ls$ as low as a few $10^{-3}$, DM particles are able to thermalize with themselves at a quite early time, and then to freeze-out after becoming non-relativistic, as explained in the four steps below. These four periods are labeled from \emph{(a)} to \emph{(d)} in the two panels of Fig.~\ref{thermal:dark}.

\begin{itemize}
\item During the period \emph{(a)}, the first DM particles are created via out-of-equilibrium decays of the Higgs ($h\to SS$).
The 2-to-4 processes are originally very suppressed by the small DM number density and the high velocity of the DM particles (initially $T'$ and $T$ are of the same order).
However these suppressions become less severe due to the increase of the DM population and to the cooling down of the dark sector.
At some point the DM production via the 2-to-4 interactions overtakes the one due to the Higgs decay, giving rise to a dramatic increase of the DM abundance at the cost of reducing its temperature.
At the end of this period DM reaches chemical equilibrium.

\begin{figure}[t]
\centering
\includegraphics[width=0.49\textwidth]{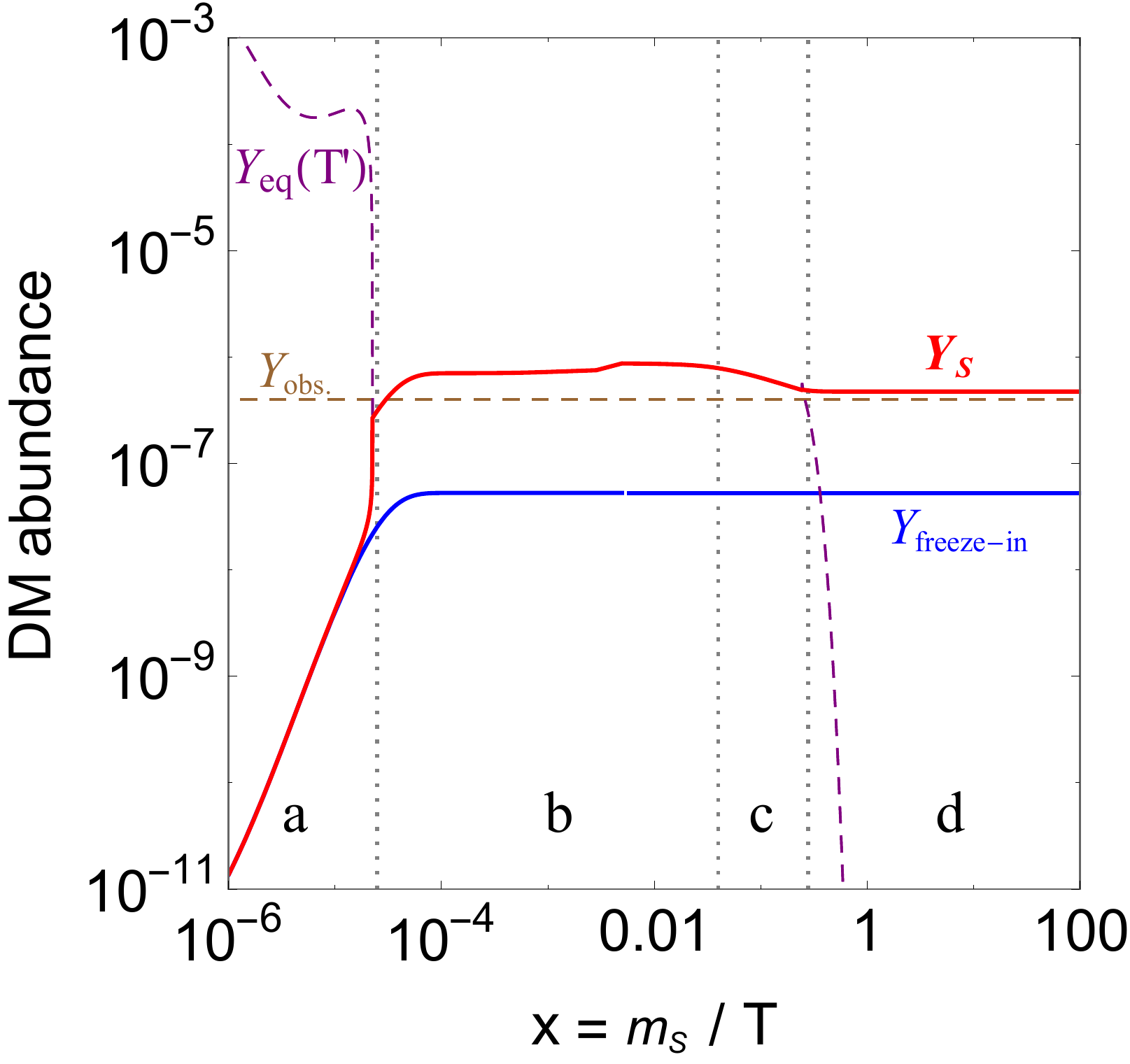}~~
\includegraphics[width=0.49\textwidth]{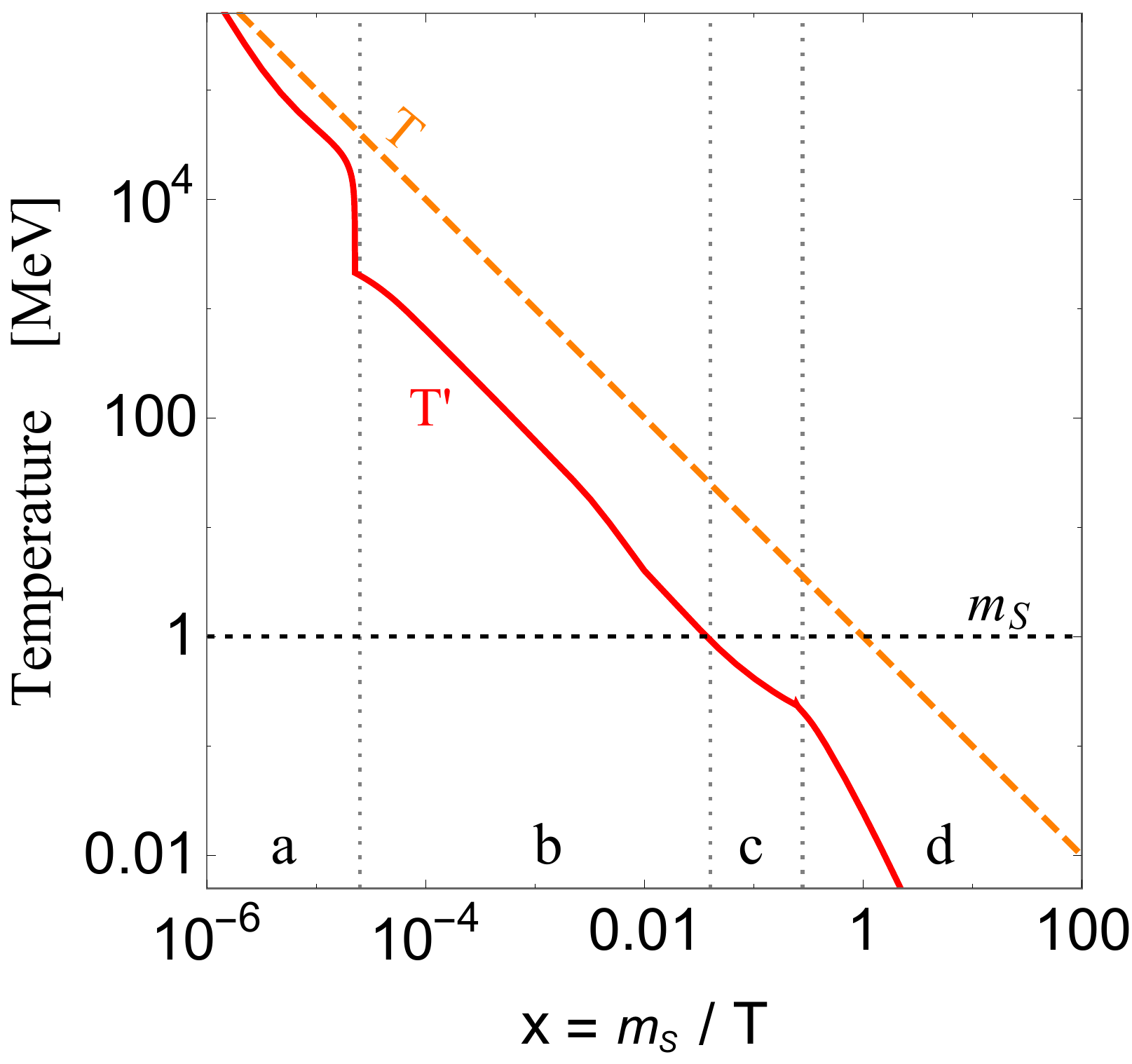}
\caption{ {\bf Left:} Evolution of DM abundance per comoving volume $Y\equiv n'/s$  with respect to the inverse photon temperature $x=\ms/T$.
$Y_S$ denotes the actual evolution of the DM density and $Y_\text{eq}(T')$ the DM equilibrium density, while $Y_\text{obs.}$ corresponds to the measured relic abundance $\Omega_\text{S}h^2\sim 0.12$~\cite{Ade:2015xua}.
$Y_\text{freeze-in}$ gives the abundance in the case where the 4-to-2 processes are ignored.
{\bf Right:} Evolution of the DM temperature $T'$ with respect to $x$.
The SM temperature $T$ and the DM mass $\ms$ (dotted horizontal line) are provided for comparison.
The parameters used are $\ms=1$~MeV, $\ls= 4\cdot 10^{-3}$ and $\lhs=6\cdot 10^{-11}$ (leading to $\xi \sim 4\cdot 10^5$). Three vertical lines stand for the times when DM reaches chemical equilibrium ($x\sim 2\cdot 10^{-5}$), DM becomes non-relativistic  ($x\sim 0.04$) and DM freezes-out ($x \sim 0.3$).
}
\label{thermal:dark}
\end{figure}

\item After reaching chemical equilibrium among themselves ($x\sim 2\cdot 10^{-5}$), in period \emph{(b)} the actual DM abundance $Y(T')$ tracks the equilibrium density $Y_\text{eq}(T')$.
Moreover, as DM is relativistic during this period, $T'$ scales as $T$.
 
\item When DM becomes non-relativistic ($T'<\ms$) in period \emph{(c)}, its number density becomes Boltzmann suppressed. During this period, due to the conservation of the DM entropy, $T'$ scales as $1/\log a$~\cite{Carlson:1992fn}, where $a$ is the scale factor. This gives rise to a relative increase of $T'$ compared to $T$.

\item Eventually when $x\sim 0.3$ (or equivalently $x'\sim 7$) DM freezes-out, leading to a constant value for $Y(T')$, which characterizes period \emph{(d)}.
During this era, DM is non-relativistic and out of chemical equilibrium, so that $T'$ decreases as $T^2$.
\end{itemize}

\begin{figure}[t]
\centering
\includegraphics[width=0.58\textwidth]{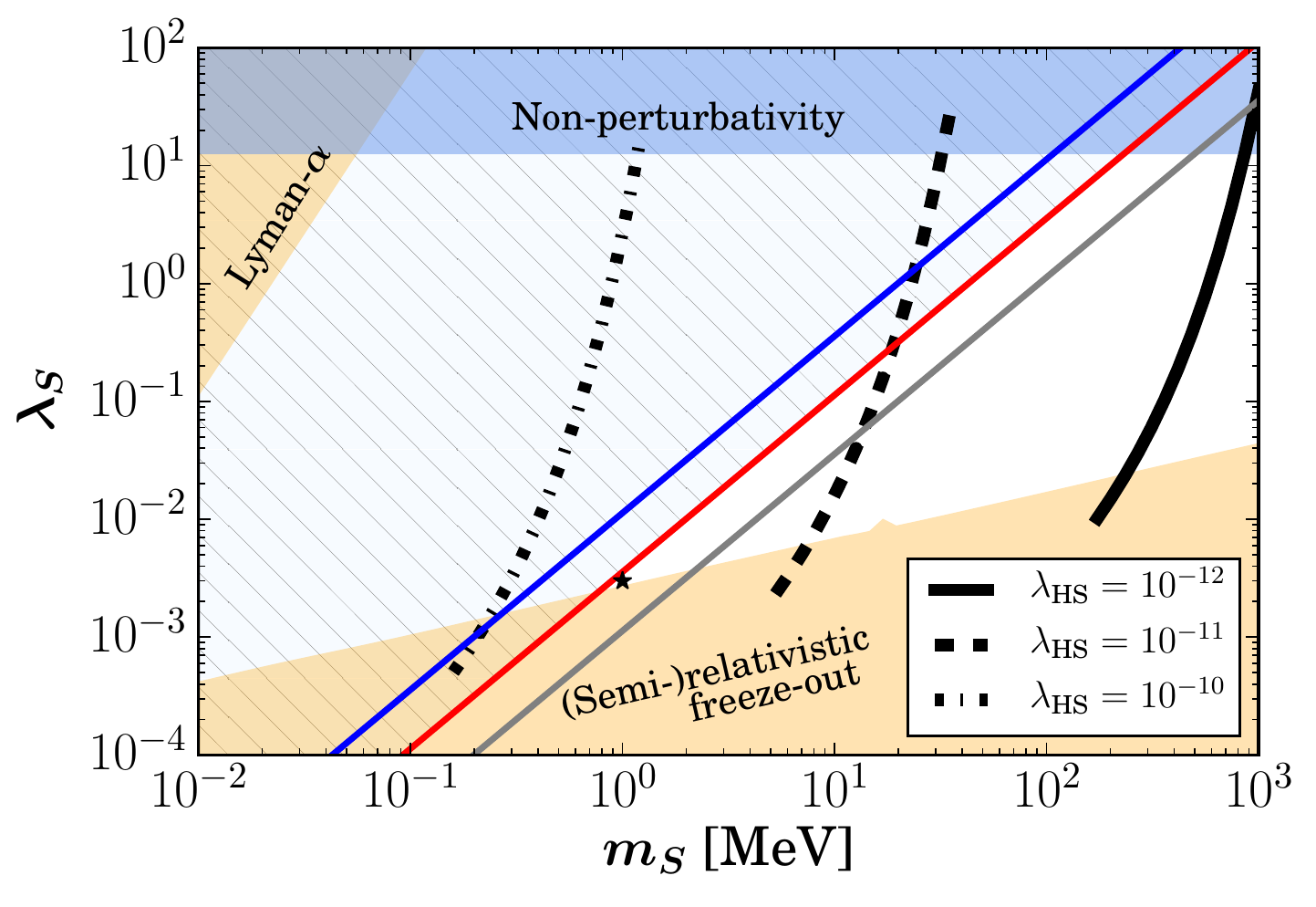}
\vspace{-.4cm}
\caption{
Same as Fig.~\ref{ms-ls}. In this case the three thick black lines correspond to different values of the Higgs portal $\lhs$ needed in order to reproduce the observed DM relic abundance.
Large values of $\lhs$ are inconsistent with DM self-interaction bounds.
}
\label{ms-ls-lhs}
\end{figure}

In the left panel of Fig.~\ref{thermal:dark}, the blue curve labeled as `freeze-in' corresponds to the case where the DM abundance is solely produced via Higgs decays (hence corresponding to the freeze-in mechanism), because the 4-to-2 processes (and then the dark thermalization) were artificially ignored.
In this case the DM relic abundance is largely underestimated.
It is worth mentioning that the importance of the 4-to-2 processes is due to the large value for $\ls$, required for the solution of the small-scale structure problems.
Therefore, in the white region of Fig.~\ref{ms-ls-lhs}, above the `(Semi-)relativistic freeze-out' band, the 4-to-2 processes have to be taken into account.
This is why the freeze-in mechanism alone can not give simultaneously a large DM self-scattering and the correct DM abundance for $\ms\gtrsim \mathcal{O}(1)$~MeV in a consistent way, in contrast to previous works which neglected the 4-to-2 processes~\cite{Campbell:2015fra,Kang:2015aqa}.

\vspace{.1cm}

Under the assumption that all the dark sector was originally created by energy-transfer from the visible sector through the Higgs portal, and taking into account that this transfer occurs mainly around $T\sim m_h/3\gg\ms$~\cite{Chu:2011be}, we calculate the ratio of energy densities between the two sectors after the Higgs decay:
\begin{equation}
\frac{\rho'}{\rho} \sim \left.\frac{\Gamma_{h\to SS}\cdot n^h_\text{eq}(T)\, m_h \cdot \frac{1}{H(T)}}{ \frac{\pi^2}{30}\,g_*^\text{SM}(T)\,T^4}\right|_{T\sim m_h/3} \sim \left.\frac{\Gamma_{h\to SS}}{H(T)\cdot g_*^\text{SM}(T) }\right|_{T\sim m_h/3}\sim 7\cdot 10^{13} \cdot\lhs^2\,,
\end{equation}
where $g_*^\text{SM}(T)$ is the effective degrees of freedom of the visible sector and $n_\text{eq}^h$ the equilibrium Higgs number density.
This expression leads to  
\begin{equation}
\xi\sim 2\cdot 10^{5} \,\cdot \,\left(\frac{10^{-10}}{\lhs}\right)^{3/2}\,,
\label{xi:lambdaHS}
\end{equation}
which is independent of $\ms$ in the relativistic limit.
Eq.~\eqref{xi:lambdaHS} together with Eq.~\eqref{entropy:observed} allow to estimate the strength of Higgs portal required to provide the observed DM relic density. This is illustrated in Fig.~\ref{ms-ls-lhs} for various values of $\lhs$. 

Besides, a DM dark freeze-out is possible before the complete decay of the Higgs population. Roughly, it corresponds to $T_\text{FO}\gtrsim m_h/3$. In this case, the (late) Higgs decay into DM particles after $T_\text{FO}$ increases the DM relic abundance, and thus resumes the DM annihilations. This is commonly referred to as the `reannihilation' mechanism~\cite{Cheung:2010gj,Cheung:2010gk,Chu:2011be}. However, we verified that this possibility never takes place in the parameter region of interest here.

\subsubsection*{Experimental Detection}
Sub-GeV DM that couples to the SM via a (very) suppressed Higgs portal is very challenging to detect, both for next generation of DM direct detection~\cite{Essig:2011nj,Essig:2012yx,Hochberg:2015pha,Chen:2015pha,Essig:2015cda} or indirect detection experiments~\cite{Essig:2013goa,Boddy:2015efa}. For instance, the experimental signature of such DM particles in direct detection searches using electron recoil events is suppressed by the smallness of both Higgs portal $\lhs$ and electronic Yukawa coupling $y_e$. 

In the context of astrophysical observations, although there exist stringent bounds on Yukawa-like couplings between SM fermions and light (meta-)stable new scalars based on stellar energy-loss arguments~\cite{Raffelt:1996wa}, they do not directly apply to $\mathbb{Z}_2$-symmetric DM particles, which can only be created in pairs. Moreover, such DM particles may be trapped in stars by strongly interacting with environmental DM particles without causing any noticeable energy loss~\cite{Jain:2005nh, Ganguly:2006ki,Zhang:2014wra}.

\section{A Caveat Leading to Lighter Dark Matter}
\label{sec:fermionicDM}

From Figs.~\ref{ms-ls} and~\ref{ms-ls-lhs}, it is clear that cluster bounds on self-interactions (together with the requirement of a non-relativistic freeze-out) impose a strong constraint that makes DM mass lie in the MeV range.
In this section, we briefly discuss scenarios where cluster bounds become less severe, opening the possibility of lighter (warm) DM freezing-out through the 4-to-2 processes.

One generic class of models which are able to avoid large self-interaction is the so-called excited (or inelastic) DM~\cite{TuckerSmith:2001hy, Finkbeiner:2007kk, Graham:2010ca, Finkbeiner:2014sja}.
Here we take as an example a pseudo-Dirac fermion invariant under a  $U(1)_D$ gauge symmetry~\cite{Finkbeiner:2007kk, Finkbeiner:2014sja, Schutz:2014nka}.
At low energies, after the symmetry breaking of the $U(1)_D$, a small Majorana mass splits the Dirac fermion into two nearly-degenerated Majorana states $\chi_1$ and $\chi_2$.
The physical spectrum contains a light state $\chi_1$, which is the DM candidate  (stabilized by an accidental $\mathbb{Z}_2$ symmetry), an excited partner $\chi_2$  slightly heavier and a massive gauge boson $A_D$. Due to the Majorana nature of the $\chi_i$, gauge interaction terms like $\chi_1\chi_1 A_D$ or $\chi_2 \chi_2 A_D$ are not allowed, while $\chi_1 \chi_2 A_D$ does exist.
Therefore, provided that $U(1)_D$ mixes with the SM hyperchange, $\chi_2$ decays to $\chi_1$ and a photon, via the kinetic mixing with $A_D$. The gauge boson $A_D$ is also unstable and decays to $\chi_1$ and $\chi_2$, or SM particles.

If $A_D$ is heavier than both $\chi_1$ and $\chi_2$, the DM abundance can not be generated via the usual freeze-out into SM particles, because its  annihilation cross section is extremely suppressed by the tiny kinetic mixing coupling (typically below $10^{-10}$ for $A_D$ mass in the MeV range~\cite{Dienes:1996zr,Redondo:2015iea}).
Nevertheless, if the gauge coupling of $U(1)_D$ is large enough,  after the decoupling of $A_D$ the chemical equilibrium among $\chi_1$ and $\chi_2$ can still be maintained via (co-)annihilation processes, such as $2\chi_i + 2\chi_j \to 2\chi_i$, $3\chi_i + \chi_j \to \chi_i+ \chi_j$ and $4\chi_i \to 2\chi_j$, with $i\ne j$.
Therefore,  the DM abundance can be set by a 4-to-2 dark freeze-out scenario.
It is similar to the one described above for the singlet scalar model, although the strength of these processes gets moderately suppressed by the existence of heavier $A_D$.

The situation of self-scatterings for excited DM, on the other hand,  differs substantially from the singlet scalar case, due to the absence of the interaction vertex $\chi_1 \chi_1 A_D$.
As a consequence, DM elastic scatterings $\chi_1 \chi_1 \to \chi_1 \chi_1$ can only happen via loops, and are thus strongly suppressed~\cite{Schutz:2014nka}.
The inelastic scattering $\chi_1 \chi_1 \to \chi_2 \chi_2$ at low temperatures is also suppressed as it is endothermal.
Therefore, cluster bounds on DM self-interactions can be easily lifted without overturning previous results, such as the dark freeze-out via 4-to-2 processes.

Such an alleviation of DM self-interaction bounds in excited DM models helps to extend the allowed parameter region. Similar conclusions can also be reached in models where DM elastic scattering is $p$-wave suppressed. Consequently, both phenomenological classes of models allow for light DM candidates, even much below MeV. In fact, for a DM mass around $\mathcal{O}$(30)~keV, the damping effect induced by its free-streaming may mimic that of conventional keV-scale warm DM on structure formation.

\section{Conclusions}
\label{sec:conclusions}

The collisionless cold dark matter (DM) paradigm has been highly successful in accounting for large scale structure of the Universe.
However, it is not clear that it can also fully explain the small scales. In fact, 
long-standing puzzles from observations of small scale objects (e.g. the `cusp vs. core' and the `too-big-to-fail' problems), can be alleviated if large DM self-interactions at scales below few Mpc exist.


The singlet scalar DM model is a very simple and economical model, described by only three parameters: the DM mass $\ms$, the DM quartic coupling $\ls$ and the Higgs portal $\lhs$.
In this framework, imposing strong DM self-interaction naturally points toward a sub-GeV $\ms$ and a self-coupling $\ls\sim\mathcal{O}(1)$. Moreover, non-observation of the invisible decay of Higgs requires a suppressed Higgs portal.
Given these conditions, it is challenging to produce the measured DM relic abundance through the standard freeze-out mechanism.
Indeed, the cross-section of DM annihilating into SM particles is very suppressed because of the smallness of both the Higgs portal and the Yukawa couplings of the relevant SM fermions.


An alternative production mechanism is to consider a dark freeze-out via $N$-to-$N'$ processes, where $N$ DM particles annihilate into $N'$ of them (with $N>N'\geq 2$), namely the SIMP paradigm.
The most studied cases correspond to 3-to-2 reactions.
However, for models where the DM stability is guaranteed by a $\mathbb{Z}_2$ symmetry, the dominant annihilation channels are 4-to-2 processes (the 3-to-2 being forbidden). Moreover, to be consistent with structure formation, DM particles colder than their SM counterparts are required before the dark freeze-out. In the singlet scalar model, it can be easily achieved by introducing a very suppressed Higgs portal.

With this in mind, we first considered the dark freeze-out of DM particles via 4-to-2 annihilations for a dark sector colder than the SM, without assuming any particular mechanism to generate such a difference of temperatures.
By analytically solving of the corresponding Boltzmann equation, we have shown that for each parameter set of ($\lambda_S$, $m_S$), there is a certain value of temperature ratio  $T/T'$, which generates both the measured DM relic abundance and the self-interactions needed to address the small-scale problems. Typically, a ratio $T/T'\sim {\mathcal O}(20)$ at the moment of the freeze-out is needed (cf. Fig.~\ref{ms-ls}).

As a second step, we studied how to generate a colder dark sector under the assumption that most of the DM particles were initially created out of the SM through the Higgs portal, à la freeze-in.
We found that for most of the parameter region considered, large values of $\ls$ coupling are able to bring DM particles into both kinetic and chemical equilibrium with themselves thanks to the fast DM elastic scatterings and the $4\leftrightarrow2$ interactions.
Therefore, in this setup one naturally obtains $T'<T$ due to the tiny portal $\lhs$.
Typical values of the Higgs portal, e.g.~$\lhs\sim\mathcal{O}(10^{-12}-10^{-10})$, needed to yield the measured DM abundance are shown in Fig.~\ref{ms-ls-lhs}  for the dark freeze-out scenario.

At last, it is worth pointing out that the essence of our conclusion is not exclusive to the singlet scalar case, but general to a specific pattern of models where DM particles solely interact with the SM via a Higgs portal:
In order to provide a possible solution to small-scale problems, such models prefer a sub-GeV DM with strong self-interactions but weakly connected to the SM.
Together with relevant experimental constraints, this challenges the usual WIMP freeze-out, but favors the dark freeze-out scenario via 4-to-2 (or 3-to-2 if allowed) annihilations of DM.
In this scenario, the measured DM relic abundance is produced by having a DM sector colder than the SM one at the moment of freeze-out, which is in turn provided by the very suppressed Higgs portal.

\acknowledgments
The authors thank Camilo Garcia-Cely and Diego Restrepo for discussions about his implementation of the model in {\tt SARAH}.
NB is supported by the São Paulo Research Foundation (FAPESP) under grants 2011/11973-4 and 2013/01792-8.


\appendix

\section{Freeze-out Approximation} \label{Boltzman:FO}
In this Appendix we closely follow the procedure presented in Ref.~\cite{Bernal:2015}.\\

After the dark and visible sectors completely decouple (i.e. the energy density transfer via the Higgs decay $h\to SS$ is not efficient anymore), the evolution of the DM number density $n'\equiv n'(T')$ is given by the Boltzmann equation:
\begin{equation}
\frac{dn'}{dt}+3H\,n'=-\langle\sigma v^3\rangle_{4\to 2}\left[n'^4-n'^2\,{n'}_\text{eq}^2\right]\,,
\label{BE}
\end{equation}
which, in the case where the SM energy density dominates the expansion of the Universe, can be rewritten in terms of the SM temperature $T$ as
\begin{equation}
x\,H(x)\frac{dY(x)}{dx} =-s(x)^3 Y(x)^2  \langle\sigma v^3\rangle_{4\to 2} \, \left[Y(x)^2-Y_\text{eq}(x)^2\right]\,,
\label{BEY}
\end{equation}
where  $x\equiv \ms/T$ and  $Y(x)\equiv n'/s(x)$.

In the non-relativistic approximation, the cross-section $\langle \sigma v^3\rangle_{4\to 2}$ is independent of the temperature, and then the solution to Eq.~\eqref{BEY} that matches the observed DM relic abundance reads
\begin{equation}
\langle \sigma v^3 \rangle_{4\to2} \sim \left( 5.8\cdot 10^{10}\, \text{GeV}^{-8}\right) \,\xf^7 \, g_{*}(T_\text{FO})^{-2.5} \left(\frac{1\,\text{GeV}}{\ms} \right)^4\,,
\label{FOcondition}
\end{equation}
Notice that the temperature that enters in this expression is the one of the SM and not the DM one.

\begin{figure}[t!]
\centering
\includegraphics[width=0.69\textwidth]{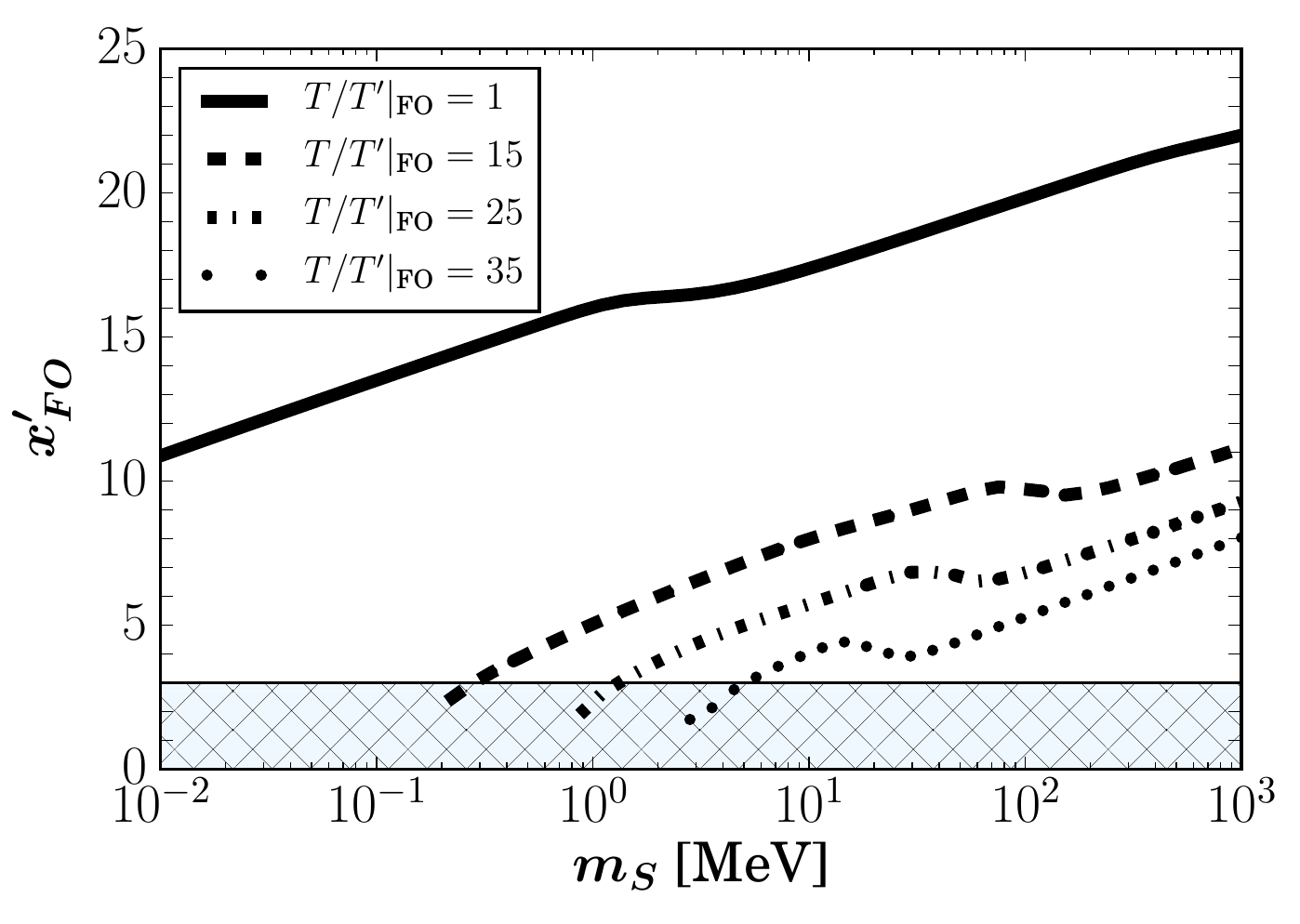}
\caption{Solution to Eq.~\eqref{xpf} for different temperature ratios at the moment of the DM freeze-out.
The hatched region corresponds to a semi-relativistic freeze-out with $x'_\text{FO}<3$.
}
\label{ms-xpfo}
\end{figure}

In order to estimate $\xf$, it is necessary to establish when  the annihilation rate per particle $(\neqq)^3 \langle \sigma v^3\rangle_{4 \to 2}$ drops below the expansion rate of the Universe. Using Eq.~\eqref{FOcondition} it is found that this happens when the freeze-out temperatures satisfy
\begin{equation}
\xpf = 17.4 + \ln \left[\left(\frac{\xf}{\xpf}\right)^3\left(\frac{\ms}{10~\text{MeV}}\right)\left(\frac{10}{g_{*\text{FO}}}\right)\left(\frac{\xpf}{17.4}\right)^{1.5}\right]\,.
\label{xpf}
\end{equation}
The solution to this equation is shown in Fig.~\ref{ms-xpfo}, for different temperature ratios at the moment of the DM freeze-out.
For $\ms=100$~keV and assuming kinetic equilibrium between the two sectors until the freeze-out, $\xf=\xpf\sim 13.5$ which agrees with the result from Ref.~\cite{Hochberg:2014dra}.

\bibliographystyle{JHEP}
\tiny
\bibliography{biblio}

\providecommand{\href}[2]{#2}\begingroup\raggedright\begin{thebibliography}{100}

\bibitem{Ade:2015xua}
{\bf Planck} Collaboration, P.~A.~R. Ade et~al., {\it {Planck 2015 results.
  XIII. Cosmological parameters}},  \href{http://arxiv.org/abs/1502.01589}{{\tt
  arXiv:1502.01589}}.

\bibitem{Flores:1994gz}
R.~A. Flores and J.~R. Primack, {\it {Observational and theoretical constraints
  on singular dark matter halos}},  {\em Astrophys.J.} {\bf 427} (1994) L1--4,
  [\href{http://arxiv.org/abs/astro-ph/9402004}{{\tt astro-ph/9402004}}].

\bibitem{Moore:1994yx}
B.~Moore, {\it {Evidence against dissipationless dark matter from observations
  of galaxy haloes}},  {\em Nature} {\bf 370} (1994) 629.

\bibitem{Oh:2010mc}
S.-H. Oh, C.~Brook, F.~Governato, E.~Brinks, L.~Mayer, et~al., {\it {The
  central slope of dark matter cores in dwarf galaxies: Simulations vs.
  THINGS}},  {\em Astron.J.} {\bf 142} (2011) 24,
  [\href{http://arxiv.org/abs/1011.2777}{{\tt arXiv:1011.2777}}].

\bibitem{Walker:2011zu}
M.~G. Walker and J.~Peñarrubia, {\it {A Method for Measuring (Slopes of) the
  Mass Profiles of Dwarf Spheroidal Galaxies}},  {\em Astrophys.J.} {\bf 742}
  (2011) 20, [\href{http://arxiv.org/abs/1108.2404}{{\tt arXiv:1108.2404}}].

\bibitem{BoylanKolchin:2011de}
M.~Boylan-Kolchin, J.~S. Bullock, and M.~Kaplinghat, {\it {Too big to fail? The
  puzzling darkness of massive Milky Way subhaloes}},  {\em
  Mon.Not.Roy.Astron.Soc.} {\bf 415} (2011) L40,
  [\href{http://arxiv.org/abs/1103.0007}{{\tt arXiv:1103.0007}}].

\bibitem{Garrison-Kimmel:2014vqa}
S.~Garrison-Kimmel, M.~Boylan-Kolchin, J.~S. Bullock, and E.~N. Kirby, {\it
  {Too Big to Fail in the Local Group}},  {\em Mon.Not.Roy.Astron.Soc.} {\bf
  444} (2014) 222, [\href{http://arxiv.org/abs/1404.5313}{{\tt
  arXiv:1404.5313}}].

\bibitem{Spergel:1999mh}
D.~N. Spergel and P.~J. Steinhardt, {\it {Observational evidence for
  self-interacting cold dark matter}},  {\em Phys.Rev.Lett.} {\bf 84} (2000)
  3760--3763, [\href{http://arxiv.org/abs/astro-ph/9909386}{{\tt
  astro-ph/9909386}}].

\bibitem{Wandelt:2000ad}
B.~D. Wandelt, R.~Dave, G.~R. Farrar, P.~C. McGuire, D.~N. Spergel, et~al.,
  {\it {Self-interacting dark matter}},
  \href{http://arxiv.org/abs/astro-ph/0006344}{{\tt astro-ph/0006344}}.

\bibitem{Vogelsberger:2012ku}
M.~Vogelsberger, J.~Zavala, and A.~Loeb, {\it {Subhaloes in Self-Interacting
  Galactic Dark Matter Haloes}},  {\em Mon.Not.Roy.Astron.Soc.} {\bf 423}
  (2012) 3740, [\href{http://arxiv.org/abs/1201.5892}{{\tt arXiv:1201.5892}}].

\bibitem{Rocha:2012jg}
M.~Rocha, A.~H. Peter, J.~S. Bullock, M.~Kaplinghat, S.~Garrison-Kimmel,
  et~al., {\it {Cosmological Simulations with Self-Interacting Dark Matter I:
  Constant Density Cores and Substructure}},  {\em Mon.Not.Roy.Astron.Soc.}
  {\bf 430} (2013) 81--104, [\href{http://arxiv.org/abs/1208.3025}{{\tt
  arXiv:1208.3025}}].

\bibitem{Peter:2012jh}
A.~H. Peter, M.~Rocha, J.~S. Bullock, and M.~Kaplinghat, {\it {Cosmological
  Simulations with Self-Interacting Dark Matter II: Halo Shapes vs.
  Observations}},  {\em Mon.Not.Roy.Astron.Soc.} {\bf 430} (2012) 105--120,
  [\href{http://arxiv.org/abs/1208.3026}{{\tt arXiv:1208.3026}}].

\bibitem{Zavala:2012us}
J.~Zavala, M.~Vogelsberger, and M.~G. Walker, {\it {Constraining
  Self-Interacting Dark Matter with the Milky Way's dwarf spheroidals}},  {\em
  Mon.Not.Roy.Astron.Soc.: Letters} {\bf 431} (2013) L20--L24,
  [\href{http://arxiv.org/abs/1211.6426}{{\tt arXiv:1211.6426}}].

\bibitem{Kaplinghat:2015aga}
M.~Kaplinghat, S.~Tulin, and H.-B. Yu, {\it {Dark Matter Halos as Particle
  Colliders: A Unified Solution to Small-Scale Structure Puzzles from Dwarfs to
  Clusters}},  \href{http://arxiv.org/abs/1508.03339}{{\tt arXiv:1508.03339}}.

\bibitem{MacLow:1998wv}
M.-M. Mac~Low and A.~Ferrara, {\it {Starburst - driven mass loss from dwarf
  galaxies: Efficiency and metal ejection}},  {\em Astrophys.J.} {\bf 513}
  (1999) 142, [\href{http://arxiv.org/abs/astro-ph/9801237}{{\tt
  astro-ph/9801237}}].

\bibitem{Governato:2009bg}
F.~Governato, C.~Brook, L.~Mayer, A.~Brooks, G.~Rhee, et~al., {\it {At the
  heart of the matter: the origin of bulgeless dwarf galaxies and Dark Matter
  cores}},  {\em Nature} {\bf 463} (2010) 203--206,
  [\href{http://arxiv.org/abs/0911.2237}{{\tt arXiv:0911.2237}}].

\bibitem{Silk:2010aw}
J.~Silk and A.~Nusser, {\it {The massive black hole-velocity dispersion
  relation and the halo baryon fraction: a case for positive AGN feedback}},
  {\em Astrophys.J.} {\bf 725} (2010) 556--560,
  [\href{http://arxiv.org/abs/1004.0857}{{\tt arXiv:1004.0857}}].

\bibitem{VeraCiro:2012na}
C.~A. Vera-Ciro, A.~Helmi, E.~Starkenburg, and M.~A. Breddels, {\it {Not too
  big, not too small: the dark halos of the dwarf spheroidals in the Milky
  Way}},  {\em Mon.Not.Roy.Astron.Soc.} {\bf 428} (2012) 1696--1703,
  [\href{http://arxiv.org/abs/1202.6061}{{\tt arXiv:1202.6061}}].

\bibitem{Massey:2015dkw}
R.~Massey, L.~Williams, R.~Smit, M.~Swinbank, T.~D. Kitching, et~al., {\it {The
  behaviour of dark matter associated with 4 bright cluster galaxies in the
  10~kpc core of Abell 3827}},  {\em Mon.Not.Roy.Astron.Soc.} {\bf 449} (2015)
  3393, [\href{http://arxiv.org/abs/1504.03388}{{\tt arXiv:1504.03388}}].

\bibitem{Kahlhoefer:2015vua}
F.~Kahlhoefer, K.~Schmidt-Hoberg, J.~Kummer, and S.~Sarkar, {\it {On the
  interpretation of dark matter self-interactions in Abell 3827}},  {\em Mon.
  Not. Roy. Astron. Soc.} {\bf 452} (2015) 1,
  [\href{http://arxiv.org/abs/1504.06576}{{\tt arXiv:1504.06576}}].

\bibitem{Ganguly:2006ki}
A.~K. Ganguly, P.~Jain, S.~Mandal, and S.~Stokes, {\it {Self Interacting Dark
  Matter in the Solar System}},  {\em Phys. Rev.} {\bf D76} (2007) 025026,
  [\href{http://arxiv.org/abs/hep-ph/0611006}{{\tt hep-ph/0611006}}].

\bibitem{Zentner:2009is}
A.~R. Zentner, {\it {High-Energy Neutrinos From Dark Matter Particle
  Self-Capture Within the Sun}},  {\em Phys. Rev.} {\bf D80} (2009) 063501,
  [\href{http://arxiv.org/abs/0907.3448}{{\tt arXiv:0907.3448}}].

\bibitem{Vogelsberger:2012sa}
M.~Vogelsberger and J.~Zavala, {\it {Direct detection of self-interacting dark
  matter}},  {\em Mon. Not. Roy. Astron. Soc.} {\bf 430} (2013) 1722--1735,
  [\href{http://arxiv.org/abs/1211.1377}{{\tt arXiv:1211.1377}}].

\bibitem{Ng:2013xha}
K.~C.~Y. Ng, R.~Laha, S.~Campbell, S.~Horiuchi, B.~Dasgupta, K.~Murase, and
  J.~F. Beacom, {\it {Resolving small-scale dark matter structures using
  multisource indirect detection}},  {\em Phys. Rev.} {\bf D89} (2014), no.~8
  083001, [\href{http://arxiv.org/abs/1310.1915}{{\tt arXiv:1310.1915}}].

\bibitem{DelNobile:2015uua}
E.~Del~Nobile, M.~Kaplinghat, and H.-B. Yu, {\it {Direct Detection Signatures
  of Self-Interacting Dark Matter with a Light Mediator}},  {\em JCAP} {\bf
  1510} (2015), no.~10 055, [\href{http://arxiv.org/abs/1507.04007}{{\tt
  arXiv:1507.04007}}].

\bibitem{Carlson:1992fn}
E.~D. Carlson, M.~E. Machacek, and L.~J. Hall, {\it {Self-interacting dark
  matter}},  {\em Astrophys.J.} {\bf 398} (1992) 43--52.

\bibitem{McDonald:2001vt}
J.~McDonald, {\it {Thermally generated gauge singlet scalars as selfinteracting
  dark matter}},  {\em Phys. Rev. Lett.} {\bf 88} (2002) 091304,
  [\href{http://arxiv.org/abs/hep-ph/0106249}{{\tt hep-ph/0106249}}].

\bibitem{Gudnason:2006yj}
S.~B. Gudnason, C.~Kouvaris, and F.~Sannino, {\it {Dark Matter from new
  Technicolor Theories}},  {\em Phys. Rev.} {\bf D74} (2006) 095008,
  [\href{http://arxiv.org/abs/hep-ph/0608055}{{\tt hep-ph/0608055}}].

\bibitem{Khlopov:2008ki}
M.~{\relax Yu}. Khlopov, {\it {Composite dark matter from stable charged
  constituents}},  \href{http://arxiv.org/abs/0806.3581}{{\tt
  arXiv:0806.3581}}.

\bibitem{Kamionkowski:2008gj}
M.~Kamionkowski and S.~Profumo, {\it {Early Annihilation and Diffuse
  Backgrounds in Models of Weakly Interacting Massive Particles in Which the
  Cross Section for Pair Annihilation Is Enhanced by 1/v}},  {\em Phys. Rev.
  Lett.} {\bf 101} (2008) 261301, [\href{http://arxiv.org/abs/0810.3233}{{\tt
  arXiv:0810.3233}}].

\bibitem{Kaplan:2009de}
D.~E. Kaplan, G.~Z. Krnjaic, K.~R. Rehermann, and C.~M. Wells, {\it {Atomic
  Dark Matter}},  {\em JCAP} {\bf 1005} (2010) 021,
  [\href{http://arxiv.org/abs/0909.0753}{{\tt arXiv:0909.0753}}].

\bibitem{Feng:2009hw}
J.~L. Feng, M.~Kaplinghat, and H.-B. Yu, {\it {Halo Shape and Relic Density
  Exclusions of Sommerfeld-Enhanced Dark Matter Explanations of Cosmic Ray
  Excesses}},  {\em Phys. Rev. Lett.} {\bf 104} (2010) 151301,
  [\href{http://arxiv.org/abs/0911.0422}{{\tt arXiv:0911.0422}}].

\bibitem{Tulin:2012wi}
S.~Tulin, H.-B. Yu, and K.~M. Zurek, {\it {Resonant Dark Forces and Small Scale
  Structure}},  {\em Phys. Rev. Lett.} {\bf 110} (2013), no.~11 111301,
  [\href{http://arxiv.org/abs/1210.0900}{{\tt arXiv:1210.0900}}].

\bibitem{Schutz:2014nka}
K.~Schutz and T.~R. Slatyer, {\it {Self-Scattering for Dark Matter with an
  Excited State}},  {\em JCAP} {\bf 1501} (2015), no.~01 021,
  [\href{http://arxiv.org/abs/1409.2867}{{\tt arXiv:1409.2867}}].

\bibitem{Hochberg:2014dra}
Y.~Hochberg, E.~Kuflik, T.~Volansky, and J.~G. Wacker, {\it {Mechanism for
  Thermal Relic Dark Matter of Strongly Interacting Massive Particles}},  {\em
  Phys.Rev.Lett.} {\bf 113} (2014) 171301,
  [\href{http://arxiv.org/abs/1402.5143}{{\tt arXiv:1402.5143}}].

\bibitem{Kouvaris:2014uoa}
C.~Kouvaris, I.~M. Shoemaker, and K.~Tuominen, {\it {Self-Interacting Dark
  Matter through the Higgs Portal}},  {\em Phys. Rev.} {\bf D91} (2015), no.~4
  043519, [\href{http://arxiv.org/abs/1411.3730}{{\tt arXiv:1411.3730}}].

\bibitem{Kusenko:2013saa}
A.~Kusenko and L.~J. Rosenberg, {\it {Working Group Report: Non-WIMP Dark
  Matter}},  in {\em {Community Summer Study 2013: Snowmass on the Mississippi
  (CSS2013) Minneapolis, MN, USA, July 29-August 6, 2013}}, 2013.
\newblock \href{http://arxiv.org/abs/1310.8642}{{\tt arXiv:1310.8642}}.

\bibitem{deLaix:1995vi}
A.~A. de~Laix, R.~J. Scherrer, and R.~K. Schaefer, {\it {Constraints of
  selfinteracting dark matter}},  {\em Astrophys.J.} {\bf 452} (1995) 495,
  [\href{http://arxiv.org/abs/astro-ph/9502087}{{\tt astro-ph/9502087}}].

\bibitem{Choi:2015bya}
S.-M. Choi and H.~M. Lee, {\it {SIMP dark matter with gauged $\mathbb{Z}_3$
  symmetry}},  {\em JHEP} {\bf 09} (2015) 063,
  [\href{http://arxiv.org/abs/1505.00960}{{\tt arXiv:1505.00960}}].

\bibitem{Bernal:2015bla}
N.~Bernal, C.~Garcia-Cely, and R.~Rosenfeld, {\it {WIMP and SIMP Dark Matter
  from the Spontaneous Breaking of a Global Group}},  {\em JCAP} {\bf 1504}
  (2015), no.~04 012, [\href{http://arxiv.org/abs/1501.01973}{{\tt
  arXiv:1501.01973}}].

\bibitem{Bernal:2015}
N.~Bernal, X.~Chu, C.~Garcia-Cely, T.~Hambye, and B.~Zaldivar, {\it {Production
  Regimes for Self-Interacting Dark Matter}},
  \href{http://arxiv.org/abs/1510.08063}{{\tt arXiv:1510.08063}}.

\bibitem{Ko:2014nha}
P.~Ko and Y.~Tang, {\it {Self-interacting scalar dark matter with local
  $\mathbb{Z}_3$ symmetry}},  {\em JCAP} {\bf 1405} (2014) 047,
  [\href{http://arxiv.org/abs/1402.6449}{{\tt arXiv:1402.6449}}].

\bibitem{Yamanaka:2014pva}
N.~Yamanaka, S.~Fujibayashi, S.~Gongyo, and H.~Iida, {\it {Dark matter in the
  hidden gauge theory}},  \href{http://arxiv.org/abs/1411.2172}{{\tt
  arXiv:1411.2172}}.

\bibitem{Hochberg:2014kqa}
Y.~Hochberg, E.~Kuflik, H.~Murayama, T.~Volansky, and J.~G. Wacker, {\it {Model
  for Thermal Relic Dark Matter of Strongly Interacting Massive Particles}},
  {\em Phys. Rev. Lett.} {\bf 115} (2015), no.~2 021301,
  [\href{http://arxiv.org/abs/1411.3727}{{\tt arXiv:1411.3727}}].

\bibitem{Lee:2015gsa}
H.~M. Lee and M.-S. Seo, {\it {Communication with SIMP dark mesons via
  Z'-portal}},  {\em Phys. Lett.} {\bf B748} (2015) 316--322,
  [\href{http://arxiv.org/abs/1504.00745}{{\tt arXiv:1504.00745}}].

\bibitem{Hansen:2015yaa}
M.~Hansen, K.~Langæble, and F.~Sannino, {\it {SIMP model at NNLO in chiral
  perturbation theory}},  {\em Phys. Rev.} {\bf D92} (2015), no.~7 075036,
  [\href{http://arxiv.org/abs/1507.01590}{{\tt arXiv:1507.01590}}].

\bibitem{McDonald:1993ex}
J.~McDonald, {\it {Gauge singlet scalars as cold dark matter}},  {\em
  Phys.Rev.} {\bf D50} (1994) 3637--3649,
  [\href{http://arxiv.org/abs/hep-ph/0702143}{{\tt hep-ph/0702143}}].

\bibitem{Burgess:2000yq}
C.~Burgess, M.~Pospelov, and T.~ter Veldhuis, {\it {The Minimal model of
  nonbaryonic dark matter: A Singlet scalar}},  {\em Nucl.Phys.} {\bf B619}
  (2001) 709--728, [\href{http://arxiv.org/abs/hep-ph/0011335}{{\tt
  hep-ph/0011335}}].

\bibitem{Barger:2007im}
V.~Barger, P.~Langacker, M.~McCaskey, M.~J. Ramsey-Musolf, and G.~Shaughnessy,
  {\it {LHC Phenomenology of an Extended Standard Model with a Real Scalar
  Singlet}},  {\em Phys. Rev.} {\bf D77} (2008) 035005,
  [\href{http://arxiv.org/abs/0706.4311}{{\tt arXiv:0706.4311}}].

\bibitem{Djouadi:2011aa}
A.~Djouadi, O.~Lebedev, Y.~Mambrini, and J.~Quevillon, {\it {Implications of
  LHC searches for Higgs--portal dark matter}},  {\em Phys. Lett.} {\bf B709}
  (2012) 65--69, [\href{http://arxiv.org/abs/1112.3299}{{\tt
  arXiv:1112.3299}}].

\bibitem{Djouadi:2012zc}
A.~Djouadi, A.~Falkowski, Y.~Mambrini, and J.~Quevillon, {\it {Direct Detection
  of Higgs-Portal Dark Matter at the LHC}},  {\em Eur. Phys. J.} {\bf C73}
  (2013), no.~6 2455, [\href{http://arxiv.org/abs/1205.3169}{{\tt
  arXiv:1205.3169}}].

\bibitem{Damgaard:2013kva}
P.~H. Damgaard, D.~O'Connell, T.~C. Petersen, and A.~Tranberg, {\it
  {Constraints on New Physics from Baryogenesis and Large Hadron Collider
  Data}},  {\em Phys. Rev. Lett.} {\bf 111} (2013), no.~22 221804,
  [\href{http://arxiv.org/abs/1305.4362}{{\tt arXiv:1305.4362}}].

\bibitem{No:2013wsa}
J.~M. No and M.~Ramsey-Musolf, {\it {Probing the Higgs Portal at the LHC
  Through Resonant di-Higgs Production}},  {\em Phys. Rev.} {\bf D89} (2014),
  no.~9 095031, [\href{http://arxiv.org/abs/1310.6035}{{\tt arXiv:1310.6035}}].

\bibitem{Robens:2015gla}
T.~Robens and T.~Stefaniak, {\it {Status of the Higgs Singlet Extension of the
  Standard Model after LHC Run 1}},  {\em Eur. Phys. J.} {\bf C75} (2015) 104,
  [\href{http://arxiv.org/abs/1501.02234}{{\tt arXiv:1501.02234}}].

\bibitem{He:2009yd}
X.-G. He, T.~Li, X.-Q. Li, J.~Tandean, and H.-C. Tsai, {\it {The Simplest
  Dark-Matter Model, CDMS II Results, and Higgs Detection at LHC}},  {\em Phys.
  Lett.} {\bf B688} (2010) 332--336,
  [\href{http://arxiv.org/abs/0912.4722}{{\tt arXiv:0912.4722}}].

\bibitem{Baek:2014jga}
S.~Baek, P.~Ko, and W.-I. Park, {\it {Invisible Higgs Decay Width vs. Dark
  Matter Direct Detection Cross Section in Higgs Portal Dark Matter Models}},
  {\em Phys. Rev.} {\bf D90} (2014), no.~5 055014,
  [\href{http://arxiv.org/abs/1405.3530}{{\tt arXiv:1405.3530}}].

\bibitem{Feng:2014vea}
L.~Feng, S.~Profumo, and L.~Ubaldi, {\it {Closing in on singlet scalar dark
  matter: LUX, invisible Higgs decays and gamma-ray lines}},  {\em JHEP} {\bf
  1503} (2015) 045, [\href{http://arxiv.org/abs/1412.1105}{{\tt
  arXiv:1412.1105}}].

\bibitem{Queiroz:2014yna}
F.~S. Queiroz and K.~Sinha, {\it {The Poker Face of the Majoron Dark Matter
  Model: LUX to keV Line}},  {\em Phys. Lett.} {\bf B735} (2014) 69--74,
  [\href{http://arxiv.org/abs/1404.1400}{{\tt arXiv:1404.1400}}].

\bibitem{Han:2015hda}
H.~Han and S.~Zheng, {\it {New Constraints on Higgs-portal Scalar Dark
  Matter}},  {\em JHEP} {\bf 12} (2015) 044,
  [\href{http://arxiv.org/abs/1509.01765}{{\tt arXiv:1509.01765}}].

\bibitem{Yaguna:2008hd}
C.~E. Yaguna, {\it {Gamma rays from the annihilation of singlet scalar dark
  matter}},  {\em JCAP} {\bf 0903} (2009) 003,
  [\href{http://arxiv.org/abs/0810.4267}{{\tt arXiv:0810.4267}}].

\bibitem{Goudelis:2009zz}
A.~Goudelis, Y.~Mambrini, and C.~Yaguna, {\it {Antimatter signals of singlet
  scalar dark matter}},  {\em JCAP} {\bf 0912} (2009) 008,
  [\href{http://arxiv.org/abs/0909.2799}{{\tt arXiv:0909.2799}}].

\bibitem{Profumo:2010kp}
S.~Profumo, L.~Ubaldi, and C.~Wainwright, {\it {Singlet Scalar Dark Matter:
  monochromatic gamma rays and metastable vacua}},  {\em Phys. Rev.} {\bf D82}
  (2010) 123514, [\href{http://arxiv.org/abs/1009.5377}{{\tt
  arXiv:1009.5377}}].

\bibitem{Cline:2013gha}
J.~M. Cline, K.~Kainulainen, P.~Scott, and C.~Weniger, {\it {Update on scalar
  singlet dark matter}},  {\em Phys.Rev.} {\bf D88} (2013) 055025,
  [\href{http://arxiv.org/abs/1306.4710}{{\tt arXiv:1306.4710}}].

\bibitem{Urbano:2014hda}
A.~Urbano and W.~Xue, {\it {Constraining the Higgs portal with antiprotons}},
  {\em JHEP} {\bf 03} (2015) 133, [\href{http://arxiv.org/abs/1412.3798}{{\tt
  arXiv:1412.3798}}].

\bibitem{Duerr:2015mva}
M.~Duerr, P.~Fileviez~Pérez, and J.~Smirnov, {\it {Scalar Singlet Dark Matter
  and Gamma Lines}},  {\em Phys. Lett.} {\bf B751} (2015) 119--122,
  [\href{http://arxiv.org/abs/1508.04418}{{\tt arXiv:1508.04418}}].

\bibitem{Duerr:2015aka}
M.~Duerr, P.~Fileviez~Pérez, and J.~Smirnov, {\it {Scalar Dark Matter: Direct
  vs. Indirect Detection}},  \href{http://arxiv.org/abs/1509.04282}{{\tt
  arXiv:1509.04282}}.

\bibitem{Hall:2009bx}
L.~J. Hall, K.~Jedamzik, J.~March-Russell, and S.~M. West, {\it {Freeze-In
  Production of FIMP Dark Matter}},  {\em JHEP} {\bf 03} (2010) 080,
  [\href{http://arxiv.org/abs/0911.1120}{{\tt arXiv:0911.1120}}].

\bibitem{Yaguna:2011qn}
C.~E. Yaguna, {\it {The Singlet Scalar as FIMP Dark Matter}},  {\em JHEP} {\bf
  08} (2011) 060, [\href{http://arxiv.org/abs/1105.1654}{{\tt
  arXiv:1105.1654}}].

\bibitem{Campbell:2015fra}
R.~Campbell, S.~Godfrey, H.~E. Logan, A.~D. Peterson, and A.~Poulin, {\it
  {Implications of the observation of dark matter self-interactions for singlet
  scalar dark matter}},  {\em Phys. Rev.} {\bf D92} (2015), no.~5 055031,
  [\href{http://arxiv.org/abs/1505.01793}{{\tt arXiv:1505.01793}}].

\bibitem{Kang:2015aqa}
Z.~Kang, {\it {View FImP miracle (by scale invariance) à la
  self-interaction}},  {\em Phys. Lett.} {\bf B751} (2015) 201--204,
  [\href{http://arxiv.org/abs/1505.06554}{{\tt arXiv:1505.06554}}].

\bibitem{Staub:2008uz}
F.~Staub, {\it {SARAH}},  \href{http://arxiv.org/abs/0806.0538}{{\tt
  arXiv:0806.0538}}.

\bibitem{Staub:2013tta}
F.~Staub, {\it {SARAH 4 : A tool for (not only SUSY) model builders}},  {\em
  Comput. Phys. Commun.} {\bf 185} (2014) 1773--1790,
  [\href{http://arxiv.org/abs/1309.7223}{{\tt arXiv:1309.7223}}].

\bibitem{Belyaev:2012qa}
A.~Belyaev, N.~D. Christensen, and A.~Pukhov, {\it {CalcHEP 3.4 for collider
  physics within and beyond the Standard Model}},  {\em Comput.Phys.Commun.}
  {\bf 184} (2013) 1729--1769, [\href{http://arxiv.org/abs/1207.6082}{{\tt
  arXiv:1207.6082}}].

\bibitem{Vogelsberger:2014pda}
M.~Vogelsberger, J.~Zavala, C.~Simpson, and A.~Jenkins, {\it {Dwarf galaxies in
  CDM and SIDM with baryons: observational probes of the nature of dark
  matter}},  {\em Mon.Not.Roy.Astron.Soc.} {\bf 444} (2014) 3684,
  [\href{http://arxiv.org/abs/1405.5216}{{\tt arXiv:1405.5216}}].

\bibitem{Elbert:2014bma}
O.~D. Elbert, J.~S. Bullock, S.~Garrison-Kimmel, M.~Rocha, J.~Oñorbe, and
  A.~H.~G. Peter, {\it {Core Formation in Dwarf Halos with Self Interacting
  Dark Matter: No Fine-Tuning Necessary}},  {\em Mon. Not. Roy. Astron. Soc.}
  {\bf 453} (2015) 29, [\href{http://arxiv.org/abs/1412.1477}{{\tt
  arXiv:1412.1477}}].

\bibitem{Markevitch:2003at}
M.~Markevitch, A.~Gonzalez, D.~Clowe, A.~Vikhlinin, L.~David, et~al., {\it
  {Direct constraints on the dark matter self-interaction cross-section from
  the merging galaxy cluster 1E0657-56}},  {\em Astrophys.J.} {\bf 606} (2004)
  819--824, [\href{http://arxiv.org/abs/astro-ph/0309303}{{\tt
  astro-ph/0309303}}].

\bibitem{Clowe:2003tk}
D.~Clowe, A.~Gonzalez, and M.~Markevitch, {\it {Weak lensing mass
  reconstruction of the interacting cluster 1E0657-558: Direct evidence for the
  existence of dark matter}},  {\em Astrophys.J.} {\bf 604} (2004) 596--603,
  [\href{http://arxiv.org/abs/astro-ph/0312273}{{\tt astro-ph/0312273}}].

\bibitem{Randall:2007ph}
S.~W. Randall, M.~Markevitch, D.~Clowe, A.~H. Gonzalez, and M.~Brada\v{c}, {\it
  {Constraints on the Self-Interaction Cross-Section of Dark Matter from
  Numerical Simulations of the Merging Galaxy Cluster 1E 0657-56}},  {\em
  Astrophys.J.} {\bf 679} (2008) 1173--1180,
  [\href{http://arxiv.org/abs/0704.0261}{{\tt arXiv:0704.0261}}].

\bibitem{Harvey:2015hha}
D.~Harvey, R.~Massey, T.~Kitching, A.~Taylor, and E.~Tittley, {\it {The
  non-gravitational interactions of dark matter in colliding galaxy clusters}},
   {\em Science} {\bf 347} (2015), no.~6229 1462--1465,
  [\href{http://arxiv.org/abs/1503.07675}{{\tt arXiv:1503.07675}}].

\bibitem{Bechtle:2014ewa}
P.~Bechtle, S.~Heinemeyer, O.~Stål, T.~Stefaniak, and G.~Weiglein, {\it
  {Probing the Standard Model with Higgs signal rates from the Tevatron, the
  LHC and a future ILC}},  {\em JHEP} {\bf 1411} (2014) 039,
  [\href{http://arxiv.org/abs/1403.1582}{{\tt arXiv:1403.1582}}].

\bibitem{Khachatryan:2014iha}
{\bf CMS} Collaboration, V.~Khachatryan et~al., {\it {Constraints on the Higgs
  boson width from off-shell production and decay to Z-boson pairs}},  {\em
  Phys.Lett.} {\bf B736} (2014) 64, [\href{http://arxiv.org/abs/1405.3455}{{\tt
  arXiv:1405.3455}}].

\bibitem{Steigman:2012nb}
G.~Steigman, B.~Dasgupta, and J.~F. Beacom, {\it {Precise Relic WIMP Abundance
  and its Impact on Searches for Dark Matter Annihilation}},  {\em Phys. Rev.}
  {\bf D86} (2012) 023506, [\href{http://arxiv.org/abs/1204.3622}{{\tt
  arXiv:1204.3622}}].

\bibitem{Feng:2008mu}
J.~L. Feng, H.~Tu, and H.-B. Yu, {\it {Thermal Relics in Hidden Sectors}},
  {\em JCAP} {\bf 0810} (2008) 043, [\href{http://arxiv.org/abs/0808.2318}{{\tt
  arXiv:0808.2318}}].

\bibitem{Ackerman:mha}
L.~Ackerman, M.~R. Buckley, S.~M. Carroll, and M.~Kamionkowski, {\it {Dark
  Matter and Dark Radiation}},  {\em Phys. Rev.} {\bf D79} (2009) 023519,
  [\href{http://arxiv.org/abs/0810.5126}{{\tt arXiv:0810.5126}}]. [,277(2008)].

\bibitem{Viel:2013apy}
M.~Viel, G.~D. Becker, J.~S. Bolton, and M.~G. Haehnelt, {\it {Warm dark matter
  as a solution to the small scale crisis: New constraints from high redshift
  Lyman-$\alpha$ forest data}},  {\em Phys. Rev.} {\bf D88} (2013) 043502,
  [\href{http://arxiv.org/abs/1306.2314}{{\tt arXiv:1306.2314}}].

\bibitem{Dev:2013yza}
P.~S. Bhupal~Dev, A.~Mazumdar, and S.~Qutub, {\it {Constraining Non-thermal and
  Thermal properties of Dark Matter}},  {\em Front. Phys.} {\bf 2} (2014) 26,
  [\href{http://arxiv.org/abs/1311.5297}{{\tt arXiv:1311.5297}}].

\bibitem{Kane:2015qea}
G.~L. Kane, P.~Kumar, B.~D. Nelson, and B.~Zheng, {\it {Dark Matter Production
  Mechanisms with a Non-Thermal Cosmological History - A Classification}},
  \href{http://arxiv.org/abs/1502.05406}{{\tt arXiv:1502.05406}}.

\bibitem{Chu:2011be}
X.~Chu, T.~Hambye, and M.~H.~G. Tytgat, {\it {The Four Basic Ways of Creating
  Dark Matter Through a Portal}},  {\em JCAP} {\bf 1205} (2012) 034,
  [\href{http://arxiv.org/abs/1112.0493}{{\tt arXiv:1112.0493}}].

\bibitem{Chu:2013jja}
X.~Chu, Y.~Mambrini, J.~Quevillon, and B.~Zaldivar, {\it {Thermal and
  non-thermal production of dark matter via Z'-portal(s)}},  {\em JCAP} {\bf
  1401} (2014), no.~01 034, [\href{http://arxiv.org/abs/1306.4677}{{\tt
  arXiv:1306.4677}}].

\bibitem{Cheung:2010gj}
C.~Cheung, G.~Elor, L.~J. Hall, and P.~Kumar, {\it {Origins of Hidden Sector
  Dark Matter I: Cosmology}},  {\em JHEP} {\bf 03} (2011) 042,
  [\href{http://arxiv.org/abs/1010.0022}{{\tt arXiv:1010.0022}}].

\bibitem{Cheung:2010gk}
C.~Cheung, G.~Elor, L.~J. Hall, and P.~Kumar, {\it {Origins of Hidden Sector
  Dark Matter II: Collider Physics}},  {\em JHEP} {\bf 03} (2011) 085,
  [\href{http://arxiv.org/abs/1010.0024}{{\tt arXiv:1010.0024}}].

\bibitem{Essig:2011nj}
R.~Essig, J.~Mardon, and T.~Volansky, {\it {Direct Detection of Sub-GeV Dark
  Matter}},  {\em Phys. Rev.} {\bf D85} (2012) 076007,
  [\href{http://arxiv.org/abs/1108.5383}{{\tt arXiv:1108.5383}}].

\bibitem{Essig:2012yx}
R.~Essig, A.~Manalaysay, J.~Mardon, P.~Sorensen, and T.~Volansky, {\it {First
  Direct Detection Limits on sub-GeV Dark Matter from XENON10}},  {\em Phys.
  Rev. Lett.} {\bf 109} (2012) 021301,
  [\href{http://arxiv.org/abs/1206.2644}{{\tt arXiv:1206.2644}}].

\bibitem{Hochberg:2015pha}
Y.~Hochberg, Y.~Zhao, and K.~M. Zurek, {\it {Superconducting Detectors for
  Super Light Dark Matter}},  \href{http://arxiv.org/abs/1504.07237}{{\tt
  arXiv:1504.07237}}.

\bibitem{Chen:2015pha}
J.-W. Chen, H.-C. Chi, C.~P. Liu, C.-L. Wu, and C.-P. Wu, {\it {Electronic and
  nuclear contributions in sub-GeV dark matter scattering: A case study with
  hydrogen}},  {\em Phys. Rev.} {\bf D92} (2015), no.~9 096013,
  [\href{http://arxiv.org/abs/1508.03508}{{\tt arXiv:1508.03508}}].

\bibitem{Essig:2015cda}
R.~Essig, M.~Fernández-Serra, J.~Mardon, A.~Soto, T.~Volansky, and T.-T. Yu,
  {\it {Direct Detection of sub-GeV Dark Matter with Semiconductor Targets}},
  \href{http://arxiv.org/abs/1509.01598}{{\tt arXiv:1509.01598}}.

\bibitem{Essig:2013goa}
R.~Essig, E.~Kuflik, S.~D. McDermott, T.~Volansky, and K.~M. Zurek, {\it
  {Constraining Light Dark Matter with Diffuse X-Ray and Gamma-Ray
  Observations}},  {\em JHEP} {\bf 11} (2013) 193,
  [\href{http://arxiv.org/abs/1309.4091}{{\tt arXiv:1309.4091}}].

\bibitem{Boddy:2015efa}
K.~K. Boddy and J.~Kumar, {\it {Indirect Detection of Dark Matter Using
  MeV-Range Gamma-Ray Telescopes}},  {\em Phys. Rev.} {\bf D92} (2015), no.~2
  023533, [\href{http://arxiv.org/abs/1504.04024}{{\tt arXiv:1504.04024}}].

\bibitem{Raffelt:1996wa}
G.~G. Raffelt, {\em {Stars as laboratories for fundamental physics}}.
\newblock 1996.

\bibitem{Jain:2005nh}
P.~Jain and S.~Mandal, {\it {Evading the astrophysical limits on light
  pseudoscalars}},  {\em Int. J. Mod. Phys.} {\bf D15} (2006) 2095--2104,
  [\href{http://arxiv.org/abs/astro-ph/0512155}{{\tt astro-ph/0512155}}].

\bibitem{Zhang:2014wra}
Y.~Zhang, {\it {Supernova Cooling in a Dark Matter Smog}},  {\em JCAP} {\bf
  1411} (2014), no.~11 042, [\href{http://arxiv.org/abs/1404.7172}{{\tt
  arXiv:1404.7172}}].

\bibitem{TuckerSmith:2001hy}
D.~Tucker-Smith and N.~Weiner, {\it {Inelastic dark matter}},  {\em Phys. Rev.}
  {\bf D64} (2001) 043502, [\href{http://arxiv.org/abs/hep-ph/0101138}{{\tt
  hep-ph/0101138}}].

\bibitem{Finkbeiner:2007kk}
D.~P. Finkbeiner and N.~Weiner, {\it {Exciting Dark Matter and the INTEGRAL/SPI
  511 keV signal}},  {\em Phys. Rev.} {\bf D76} (2007) 083519,
  [\href{http://arxiv.org/abs/astro-ph/0702587}{{\tt astro-ph/0702587}}].

\bibitem{Graham:2010ca}
P.~W. Graham, R.~Harnik, S.~Rajendran, and P.~Saraswat, {\it {Exothermic Dark
  Matter}},  {\em Phys. Rev.} {\bf D82} (2010) 063512,
  [\href{http://arxiv.org/abs/1004.0937}{{\tt arXiv:1004.0937}}].

\bibitem{Finkbeiner:2014sja}
D.~P. Finkbeiner and N.~Weiner, {\it {An X-Ray Line from eXciting Dark
  Matter}},  \href{http://arxiv.org/abs/1402.6671}{{\tt arXiv:1402.6671}}.

\bibitem{Dienes:1996zr}
K.~R. Dienes, C.~F. Kolda, and J.~March-Russell, {\it {Kinetic mixing and the
  supersymmetric gauge hierarchy}},  {\em Nucl. Phys.} {\bf B492} (1997)
  104--118, [\href{http://arxiv.org/abs/hep-ph/9610479}{{\tt hep-ph/9610479}}].

\bibitem{Redondo:2015iea}
J.~Redondo, {\it {Atlas of solar hidden photon emission}},  {\em JCAP} {\bf
  1507} (2015), no.~07 024, [\href{http://arxiv.org/abs/1501.07292}{{\tt
  arXiv:1501.07292}}].

\end{thebibliography}\endgroup

\end{document}